\documentclass[prd,eqsecnum,twocolumn,amsfonts,amssymb]{revtex4}

\usepackage{graphicx}

\usepackage{bm}

\newcommand{\beq}{\begin{equation}}
\newcommand{\eeq}{\end{equation}}
\newcommand{\beqs}{\begin{eqnarray}}
\newcommand{\eeqs}{\end{eqnarray}}

\newcommand{\drawsquare}[2]{\hbox{%
\rule{#2pt}{#1pt}\hskip-#2pt%  left vertical
\rule{#1pt}{#2pt}\hskip-#1pt%  lower horizontal
\rule[#1pt]{#1pt}{#2pt}}\rule[#1pt]{#2pt}{#2pt}\hskip-#2pt%  upper horizontal
\rule{#2pt}{#1pt}}% right vertical
\newcommand{\fund}{\raisebox{-.5pt}{\drawsquare{6.5}{0.4}}}%  fund
\newcommand{\sym}{\raisebox{-.5pt}{\drawsquare{6.5}{0.4}}\hskip-0.4pt%
        \raisebox{-.5pt}{\drawsquare{6.5}{0.4}}}%  symmetric second rank
\newcommand{\asym}{\raisebox{-3.5pt}{\drawsquare{6.5}{0.4}}\hskip-6.9pt%
        \raisebox{3pt}{\drawsquare{6.5}{0.4}}}%  antisymmetric second rank

\begin{document}

\title{Higher-Loop Structural Properties of the $\beta$ Function in
 Asymptotically Free \\ Vectorial Gauge Theories} 

\author{Robert Shrock\footnote{On leave from 
C. N. Yang Institute for Theoretical Physics, 
Stony Brook University, Stony Brook, NY 11794}}

\affiliation{Department of Physics, Sloane Laboratory \\
Yale University, New Haven, CT 06520} 

\begin{abstract}

We investigate some higher-loop structural properties of the $\beta$ function
in asymptotically free vectorial gauge theories. Our main focus is on theories
with fermion contents that lead to an infrared (IR) zero in $\beta$.  We
present analytic and numerical calculations of the value of the gauge coupling
where $\beta$ reaches a minimum, the value of $\beta$ at this minimum, and the
slope of $\beta$ at the IR zero, at two-, three-, and four-loop order. The
slope of $\beta$ at the IR zero is relevant for estimates of a dilaton mass in
quasiconformal gauge theories. Some inequalities are derived concerning the
dependence of the above quantities on loop order.  A general inequality is
derived concerning the dependence of the shift of the IR zero of $\beta$, from
the $n$-loop to the $(n+1)$-loop order, on the sign of the $(n+1)$-loop
coefficient in $\beta$.  Some results are also given for gauge theories with
${\cal N}=1$ supersymmetry.

\end{abstract}


\maketitle

\section{Introduction}

The evolution of an asymptotically free gauge theory from high Euclidean
momentum scales $\mu$ in the deep ultraviolet (UV) to small scales in the
infrared is of fundamental field-theoretic interest.  This evolution is
described by the $\beta$ function of the theory. Following the pioneering
calculations of the $\beta$ function at one-loop \cite{b1} and two-loop
\cite{b2} order, this function was subsequently calculated to three-loop
\cite{b3} and four-loop \cite{b4} order in the modified minimal \cite{ms}
subtraction ($\overline{MS}$) scheme \cite{msbar}. The anomalous dimension of
the (gauge-invariant) fermion bilinear operator, $\gamma_m$, has also been
calculated up to four-loop order in this scheme \cite{gamma4}.

Here we consider the UV to IR evolution of an asymptotically free vectorial
gauge theory with gauge group $G$ and $N_f$ massless fermions transforming
according to a representation $R$ of $G$ \cite{mf}. An interesting property of
this type of theory is that, for sufficiently large $N_f$, the two-loop $\beta$
function has an IR zero \cite{b2,bz}. If $N_f$ is near to the maximum allowed
by the property of asymptotic freedom, then this IR zero occurs at a small
value, but, as $N_f$ decreases, it increases to stronger coupling. This
motivates the calculation of the IR zero of $\beta$ at higher-loop order
\cite{gk98}.  Calculations of this IR zero, and the associated anomalous
dimension of the (gauge-invariant) fermion bilinear, $\gamma_m$, have recently
been done to four-loop order for an asymptotically free vectorial gauge theory
with gauge group $G$ and $N_f$ fermions in an arbitrary representation $R$,
with explicit results for $R$ equal to the fundamental, adjoint, and symmetric
and antisymmetric rank-2 tensor representations \cite{bvh,ps}.  A corresponding
analysis was carried out for an asymptotically free vectorial gauge theory with
${\cal N}=1$ supersymmetry in \cite{bfs}. Although the terms in the $\beta$
function at three- and higher-loop order, and the terms in $\gamma_m$ at two-
and higher-loop order are dependent on the scheme used for regularization and
renormalization of the theory, these higher-loop calculations are valuable
because they give a quantitative measure of the accuracy and stability of the
lowest-order calculations of $\alpha_{IR}$ and $\gamma_m$.  A study of the
effect of scheme transformations on results for $\alpha_{IR}$ was performed in
\cite{sch}.

In this paper we will present calculations at the $n$-loop level, where $n=2, \
3, \ 4$, of several important quantities that provide a detailed description of
the UV to IR evolution of a theory with an IR zero in its $\beta$ function.
Our general results apply for an arbitrary (non-Abelian) gauge group $G$. We
denote the running gauge coupling at a scale $\mu$ as $g(\mu)$, and define
$\alpha(\mu)=g(\mu)^2/(4\pi)$. (The $\mu$ argument will often be suppressed in
the notation.) The loop order to which a quantity is calculated is indicated
explicitly via the subscript $n\ell$, standing for $n$-loop, so that the
$n$-loop $\beta$ function and its IR zero are denoted $\beta_{n\ell}$ and
$\alpha_{IR,n\ell}$.  Given the asymptotic freedom of the theory, the UV to IR
evolution, as described by $\beta_{n\ell}$, occurs in the interval
\beq
I_\alpha: \quad 0 \le \alpha(\mu) \le \alpha_{IR,n\ell} \ . 
\label{alphainterval}
\eeq
In addition to $\alpha_{IR,n\ell}$, the three structural properties of
$\beta_{n\ell}$ that we study are (i) the value of $\alpha$ where
$\beta_{n\ell}$ reaches its minimum in the interval (\ref{alphainterval}),
denoted $\alpha_{m,n\ell}$, (ii) the minimum value of $\beta_{n\ell}$ on this
interval, $(\beta_{n\ell})_{min}$, and (iii) the slope of $\beta_{n\ell}$ at
$\alpha_{IR,n\ell}$, denoted $d\beta_{n\ell}/d\alpha {}|_{\alpha_{IR,n\ell}}$.
The importance of the first two quantities for the UV to IR evolution of the
theory is clear. One would like to know where the rate of running,
$\beta=d\alpha/dt$, has maximum magnitude, as a function of $\alpha$, and
hence, as a function of $\mu$.  Further, one is interested in what this maximum
magnitude in the rate of running, i.e., (since $\beta \le 0$), the minimum
value of $\beta$ is in the interval $I_\alpha$.  The third quantity, the slope
of the $\beta$ function at $\alpha_{IR}$, is of interest because it describes
how rapidly $\beta$ approaches zero as $\alpha$ approaches $\alpha_{IR}$. A
knowledge of this slope is also valuable because it is relevant for estimates
of a dilaton mass in gauge theories that exhibit approximate scale-invariance
associated with an IR zero of $\beta$ at a value, $\alpha_{IR}$, that is
sufficiently large that this approximate dilatation symmetry is broken by the
formation of a fermion condensate.  For each of these structural quantities,
one would like to see how higher-loop calculations compare with the two-loop
computation.  As part of our work, we derive some inequalities concerning the
relative values of each of these quantities at the two- and three-loop order.

We also generalize some results that were obtained in \cite{bvh} concerning
$\alpha_{IR,n\ell}$. In \cite{bvh} it was shown that in a theory with a given
$G$, $R$, and $N_f$ for which the two-loop $\beta$ function $\beta_{2\ell}$ has
an IR zero, the three-loop zero satisfies the inequality $\alpha_{IR,3\ell} <
\alpha_{IR,2\ell}$ in the minimal subtraction ($\overline{MS}$) scheme used
there \cite{alfdif34}.  The reduction in the value of the IR zero going from
two-loop to three-loop level is typically substantial; for example, for $G={\rm
SU}(2)$ and $N_f=8$, $\alpha_{IR,2\ell}=1.26$, while $\alpha_{IR,3\ell}=0.688$,
and for $G={\rm SU}(3)$ and $N_f=12$, $\alpha_{IR,2\ell}=0.754$, while
$\alpha_{IR,3\ell}=0.435$.  A natural question that arises from the analysis in
\cite{bvh} is how general this inequality is and, specifically, whether it also
holds for other schemes. We address and answer this question here.  We prove
that for an asymptotically free theory with a given $G$, $R$, and $N_f$ for
which $\beta_{2\ell}$ has an IR zero, the inequality $\alpha_{IR,3\ell} <
\alpha_{IR,2\ell}$ holds in any scheme which has the property that the sign of
the three-loop coefficient in $\beta$ is opposite to that of the one-loop
coefficient for $N_f \in I$, which thus preserves for $\beta_{3\ell}$ the
existence of an IR zero that was true of $\beta_{2\ell}$. This preservation of
the two-loop IR zero in $\beta$ is physically desirable, since $\beta_{2\ell}$
is scheme-independent, so if it exhibits an IR zero, then a reasonable scheme
should maintain the existence of this zero at higher-loop level. More
generally, we will derive a result that shows how $\alpha_{IR,n\ell}$ shifts,
upward or downward, to $\alpha_{IR,(n+1)\ell}$, when it is calculated to the
next higher-loop order.

For a given gauge group $G$, the infrared properties of the theory depend on
the fermion representation $R$ and the number of fermions, $N_f$.  For a
sufficiently large number, $N_f$, of fermions in a given representation (as
bounded above by the requirement of asymptotic freedom), the IR zero in $\beta$
occurs at a relatively small value of $\alpha$ and the theory evolves from the
UV to the IR without any spontaneous chiral symmetry breaking (S$\chi$SB). In
this case, the IR zero of $\beta$ is an exact infrared fixed point of the
renormalization group. Thus, the infrared behavior of the theory exhibits
scale-invariance (actually conformal invariance \cite{dilconf}) in a
non-Abelian Coulomb phase.  For small $N_f$, as the theory evolves from the UV
to the IR, and the reference scale $\mu$ decreases below a scale which may be
denoted $\Lambda$, the gauge interaction becomes strong enough to confine and
produce bilinear fermion condensates, with the associated spontaneous chiral
symmetry breaking and dynamical generation of fermion masses of order
$\Lambda$.  As $\mu$ decreases below $\Lambda$, and one constructs the
effective low-energy field theory applicable in this region, one thus
integrates out these now-massive fermions, and the $\beta$ function changes to
that of a pure gauge theory, which does not have any perturbative IR zero.
Hence, in this case the infrared zero of $\beta$ is an approximate, but not
exact, fixed point of the renormalization group.

If $N_f$ is only slightly less than the critical value $N_{f,cr}$ for
spontaneous chiral symmetry breaking, so that $\alpha_{IR}$ is only slightly
greater than the critical value, $\alpha_{cr}$ (depending on $G$ and $R$) for
fermion condensation, then the UV to IR evolution exhibits approximate scale
(dilatation) invariance for an extended logarithmic interval, because as
$\alpha(\mu)$ increases toward $\alpha_{IR}$, while less than $\alpha_{cr}$,
$\beta$ approaches zero, i.e., the rate of change of $\alpha(\mu)$ as a
function of $\mu$ approaches zero.  Thus, $\alpha(\mu)$ is large, of O(1), but
slowly running (``walking'').  This is quite different from the behavior of
$\alpha_s(\mu)$ in quantum chromodynamics (QCD).  This approximate scale
invariance at strong coupling plays an important role in models with dynamical
electroweak symmetry breaking \cite{otherwtc,chipt}, and occurs naturally in
models with an approximate infrared fixed point \cite{chipt}.  Since
$\alpha_{IR} \sim O(1)$ and $\gamma_m$ is a power series in $\alpha$, there is
an enhancement of $\gamma_m$ in such models, which, in turn, is useful for
generating sufficiently large Standard Model fermion masses. Approximate
calculations of hadron masses and related quantities have been performed using
continuum field theoretic methods for these theories \cite{sg}. Recently, an
intensive effort has been made using lattice methods to study the properties of
SU($N_c$) gauge theories with various fermion contents, in particular, theories
that exhibit quasi-scale-invariant behavior associated witn an exact or
approximate IR zero of the respective $\beta$ functions. For example, for SU(3)
with fermions in the fundamental representation, measurements of $\gamma_m$
have been reported in \cite{lgt}. In theories where $N_c$, $R$, and $N_f$ are
such that $\alpha_{IR}$ is only slightly greater than $\alpha_{cr}$, so this
approximate scale invariance associated with an IR zero of $\beta$ at strong
coupling holds, the spontaneous breaking of this symmetry by the formation of a
fermion condensate, may lead to a light state which is an approximate
Nambu-Goldstone boson (NGB), the dilaton \cite{dil} (see also \cite{sg}).  The
mass of the dilaton depends on several quantities, including the effective
value of the $\beta$ function at the relevant scale $\mu \sim \Lambda$ where
the S$\chi$SB takes place.  The desire to study the quasi-scale-invariant
behavior of such a theory is an important motivation for obtaining more
detailed information about the structure of the $\beta$ function, as contained
in the structural quantities (i)-(iii) discussed above.

% ========================================================================

\section{Beta Function} 
\label{betafunction}

\subsection{General} 

The UV to IR evolution of the theory is described by the $\beta$ 
function
\beq
\beta \equiv \beta_\alpha \equiv \frac{d\alpha}{dt} \ , 
\label{betadef}
\eeq
where $t=\ln \mu$. This has the series expansion
\beq
\beta = -2\alpha \sum_{\ell=1}^\infty b_\ell \, a^\ell =
 -2\alpha \sum_{\ell=1}^\infty \bar b_\ell \, \alpha^\ell \ ,
\label{beta}
\eeq
where $\ell$ denotes the number of loops involved in the calculation of
$b_\ell$, $a \equiv g^2/(16\pi^2) = \alpha/(4\pi)$, and 
\beq
\bar b_\ell = \frac{b_\ell}{(4\pi)^\ell} \ .
\label{bellbar}
\eeq
As noted above, the one- and two-loop coefficients $b_1$ and $b_2$, which are
scheme-independent, were calculated in \cite{b1,b2} (see Appendix A).  The
$b_\ell$ for $\ell \ge 3$ are scheme-dependent; in the commonly used
$\overline{MS}$ scheme, the $b_\ell$ have been calculated up to four-loop order
\cite{b3,b4}.  For analytical purposes it is more convenient to deal with the
$b_\ell$, since they are free of factors of $4 \pi$. However, for numerical
purposes, it is usually more convenient to use the $\bar b_\ell$, since, as is
evident from Table I of \cite{bvh}, the range of values of $\bar b_\ell$ is
smaller than the range for the $b_\ell$. We
will use both interchangeably.  We denote the $\beta$ function calculated to
$n$-loop order as
\beq
\beta_{n\ell} = - 8\pi \sum_{\ell=1}^n b_\ell \, a^{\ell+1} = 
-2\sum_{\ell=1}^n \bar b_\ell \, \alpha^{\ell+1} \ . 
\label{betanell}
\eeq
Some explicit examples of four-loop $\beta$ functions are given in 
Appendix B. 

With our sign conventions, the restriction to an asymptotically free theory
means that $b_1 > 0$.  This is equivalent to the condition 
\beq
N_f < N_{f,b1z} \ , 
\label{nfrange}
\eeq
where \cite{casimir,nfintegral} 
\beq
N_{f,b1z} = \frac{11C_A}{4T_f} \ . 
\label{nfb1z} 
\eeq
($b \ell z$ stands for $b_\ell$ zero). For the fundamental, adjoint, and
symmetric and antisymmetric rank-2 tensor representations of $G={\rm SU}(N_c)$,
the upper bound (\ref{nfrange}) allows the following ranges of $N_f$: (i) $N_f
< (11/2)N_c$ for fundamental, (ii) $N_f < 11/4$ for adjoint, (iii) $N_f <
11N_c/[2(N_c \pm 2)]$ for symmetric (antisymmetric) rank-2 tensor.  In the case
of a sufficiently large representation $R$, this upper bound may forbid even
the value $N_f=1$.  For example, for the rank-3 symmetric tensor representation
of ${\rm SU}(N_c)$, the upper bound is $N_f < 11N_c/[(N_c+2)(N_c+3)]$, and the
right-hand side of this bound is larger than 1 only for $N_c$ (analytically
continued to nonnegative real numbers) in the interval $3-\sqrt{3} < N_f <
3+\sqrt{3}$, i.e., $1.268 < N_f < 4.732$ (to the indicated floating-point
accuracy).  Hence, if $N_c$ is equal to 2, 3, or 4, the bound allows only the
single value $N_f=1$, and if $N_c \ge 5$, then the bound does not allow any
nonzero (integer) value of $N_f$.  For $G={\rm SU}(2)$, with a representation
labeled by the integer or half-integer $j$, the inequality (\ref{nfrange}) is
\beq
N_f < \frac{33}{2j(j+1)(2j+1)}  \quad {\rm for} \ \ G = {\rm SU}(2) \ . 
\label{nfrange_su2}
\eeq
This bound is: (i) $N_f < 11$ if $j=1/2$; (ii) $N_f< 11/4$ if $j=1$; (iii) $N_f
< 11/10$ if $j=3/2$.  The right-hand side of (\ref{nfrange_su2}) decreases
through 1 as $j$ (continued to real numbers) increases through 1.562, so that
the upper bound (\ref{nfrange_su2}) does not allow a nonzero number of fermions
in a representation of SU(2) with $j \ge 2$ \cite{othergg}. 

To analyze the zeros of the $n$-loop $\beta$ function, $\beta_{n\ell}$, aside
from the double zero at $\alpha=0$, one extracts the overall factor of 
$-2\alpha^2$ and calculates the zeros of the reduced $(r)$ polynomial
\beq
\beta_{n\ell,r} \equiv 
-\frac{\beta_{n\ell}}{2\alpha^2} = \sum_{\ell=1}^n \bar b_\ell \, 
\alpha^{\ell-1} \ . 
\label{beta_nloop_factor}
\eeq
or equivalently, $ \sum_{\ell=1}^n b_\ell \, a^{\ell-1}$.  As is clear from
Eq. (\ref{beta_nloop_factor}), the zeros of $\beta_{n\ell}$ away from the
origin depend only on $n-1$ ratios of coefficients, which can be taken as 
$\bar b_\ell/\bar b_n$ for $\ell=1,...,n-1$.  Although
Eq. (\ref{beta_nloop_factor}) is an algebraic equation of degree $n-1$, with
$n-1$ roots, only one of these is physically relevant as the IR zero of
$\beta_{n\ell}$.  We denote this as $\alpha_{IR,n\ell}$.  In analyzing how the
$n$-loop $\beta$ function describes the UV to IR evolution of the theory, we
will focus on the interval (\ref{alphainterval}).  

To investigate how $\alpha_{IR,n\ell}$ changes when one calculates it to
higher-loop order, it is useful to characterize the full set of zeros of
$\beta_{n\ell}$.  In general, if one has a polynomial of degree $m$, $P_m(z) =
\sum_{s=0}^m \kappa_s z^s$, and one denotes the set of $m$ roots of the
equation $P_m(z)=0$ as $\{z_1,...,z_m \}$, then the discriminant of this
equation is defined as \cite{disc}
\beq
\Delta_m \equiv \Big [ \kappa_m^{m-1} \prod_{i < j} (z_i-z_j) \Big ]^2 \ . 
\label{deltam}
\eeq

Since $\Delta_m$ is a symmetric polynomial in the roots of the equation
$P_m(z)=0$, the symmetric function theorem implies that it can be expressed as
a polynomial in the coefficients of $P_m(z)$ \cite{symfun}. We will sometimes
indicate this dependence explicitly, writing $\Delta_m(\kappa_0,...,\kappa_m)$.
The discriminant $\Delta_m$ is a homogeneous polynomial of degree $m(m-1)$ in
the roots $\{ z_i \}$.  For our present purpose, to analyze the zeros of
$\beta_{n\ell}$ away from the origin, given by the roots of
Eq. (\ref{beta_nloop_factor}), of degree $m=n-1$, we will thus use the
discriminant $\Delta_{n-1}(\bar b_1, \bar b_2,...,\bar b_n)$, or equivalently,
$\Delta_{n-1}(b_1,b_2,...,b_n)$. Note that, because of the homogeneity
properties,
\beq
\Delta_{n-1}(\bar b_1,\bar b_2,...,\bar b_n) = 
(4\pi)^{-(n+1)(n-2)}\Delta_{n-1}(b_1,b_2,...,b_n) \ .
\label{deltambbar}
\eeq
Some further details on discriminants are given in Appendix C.

Although we focus on the behavior of $\beta_{n\ell}$ in the physical interval
(\ref{alphainterval}), in characterizing the zeros of $\beta_{n\ell}$, we will
make use of some formal mathematical properties of $\beta_{n\ell}$ as an
abstract function of $\alpha$. For large $|\alpha|$, since $\beta_{n\ell} \sim
-2\bar b_n \alpha^{n+1}$, it follows that if $\alpha$ is large and positive,
then ${\rm sgn}(\beta_{n\ell}) = -{\rm sgn}(b_n)$, while for large negative
$\alpha$, ${\rm sgn}(\beta_{n\ell}) = {\rm sgn}((-1)^n b_n)$.  Thus, ${\rm
sgn}(\beta_{n\ell})$ for large positive $\alpha$ is equal to $(-1)^{n+1}{\rm
sgn}(\beta_{n\ell})$ for large negative $\alpha$. Since $\beta_{n\ell}$ is
negative in the vicinity of the origin, it follows that for (both even and odd)
$n \ge 2$,
\beqs
& & {\rm If} \ b_n < 0, \ {\rm then} \ \beta_{n\ell} \ {\rm has \ at \ least \
one \ zero \ at \ a} \cr\cr
& & {\rm positive \ real \ value \ of } \ \alpha \ .
\label{bnneg}
\eeqs
Furthermore, again because $\beta \sim -2\bar b_n \alpha^{n+1}$ for large
$|\alpha|$, a consequence is that for $n \ge 2$, 
\beqs
& & {\rm If} \ n \ {\rm is \ odd \ and} \ b_n < 0 \ {\rm or} \ n \ 
{\rm is \ even \ and} \ b_n > 0, \ {\rm then} \cr\cr
         & & \beta_{n\ell} \ {\rm has \ at \ least \ one \ zero \ at \ a \
  negative \ real} \cr\cr
& & {\rm value \ of} \ \alpha \ .
\label{bnpos}
\eeqs
Of course, the behavior of $\beta_{n\ell}$ at negative values of $\alpha$ is
not directly physical, and the behavior at large positive $\alpha$ is beyond
the range of validity of the perturbative calculation, but these mathematical
properties will be useful in characterizing the total set of zeros of
$\beta_{n\ell}$ at higher-loop order.

Given that $N_f \in I$, so that $\beta_{2\ell}$ has an IR zero, we can track
how this zero changes as the loop order $n$ increases.  One
general result is as follows. As $N_f \nearrow
N_{f,b1z}$ at the upper end of the interval $I$, $\alpha_{IR,n\ell} \to
0$. This is a result of the fact that in this limit, $\bar b_1 \to 0$,
so that $\beta_{n\ell,r}$ reduces to 
$\alpha \sum_{\ell=2}^n \alpha^{\ell-2}$, which has a root at $\alpha=0$. 
Starting at the $(n=2)$-loop level and tracking the physical IR zero at three-
and higher-loop order, one can infer that generically $\alpha_{n,\ell}$ is the
root of $\beta_{n\ell,r}$ that moves toward zero in this limit $N_f \nearrow
N_{f,b1z}$. 

Because $\beta_{n\ell}$ is a polynomial in $\alpha(\mu)$ and hence a continuous
function, and because $\beta_{n\ell}=0$ at the two ends of the interval
(\ref{alphainterval}), at $\alpha=0$ and $\alpha=\alpha_{IR,n\ell}$, and is
negative for small (positive) $\alpha$, it follows that $\beta$ reaches a
minimum in this interval (\ref{alphainterval}).  This occurs at a point where 
$d\beta_{n\ell}/d\alpha=0$, which we label $\alpha_{m,n\ell}$ (where the
subscript $m$ stands for ``minimum $\beta$ in $I_\alpha$''), and
we denote
\beq
(\beta_{n\ell})_{min} \equiv \beta_{n\ell}{}\Big |_{\alpha=\alpha_{m,n\ell}} 
\ . 
\label{betamin_n-loop}
\eeq

From Eq. (\ref{betanell}), one calculates 
$d\beta_{n\ell}/d\alpha = (4\pi)^{-1} d\beta_{n\ell}/da$, with the result 
\beq
\frac{d\beta_{n\ell}}{d\alpha}=-2 \sum_{\ell=1}^n (\ell+1) b_\ell \, a^\ell = 
-2 \sum_{\ell=1}^n (\ell+1) \bar b_\ell \, \alpha^\ell \ . 
\label{dbetanell_dalpha}
\eeq
The equation for the critical points, where $d\beta_{n\ell}/d\alpha=0$, is thus
an algebraic equation of degree $n$, with $n$ formal roots, one of which is
$\alpha=0$.  Assuming that $b_2 < 0$, i.e., $N_f \in I$, so that the two-loop
$\beta$ function has an IR zero (and also, in higher-loop calculations, that
the scheme preserves the existence of this IR zero), it follows that, among the
remaining $n-1$ roots, one is real and positive and yields the minimum value of
$\beta_{n\ell}$ for $\alpha$ in the relevant interval (\ref{alphainterval}),
and this root is the above-mentioned $\alpha_{m,n\ell}$.

Given that $\beta$ has an IR zero at $\alpha_{IR}$ and is analytic at this
point, one may expand it in a Taylor series about $\alpha_{IR}$. This involves
the slope of the $\beta$ function at $\alpha_{IR}$.  For compact notation, we
denote
\beq
\beta'_{IR} \equiv \frac{d\beta}{d\alpha} {}\Big |_{\alpha=\alpha_{IR}} 
\label{beta'}
\eeq
and, for the $n$-loop quantities,
\beq
\beta'_{IR,n\ell} \equiv \frac{d\beta_{n\ell}}{d\alpha} 
{}\Big |_{\alpha=\alpha_{IR,n\ell}} \ . 
\label{beta'_nloop}
\eeq
With $\beta(\alpha_{IR})=0$, the expansion of $\beta(\alpha)$ for $\alpha$ near
to $\alpha_{IR}$ is 
\beq
\beta = \beta'_{IR} \, (\alpha-\alpha_{IR}) + 
O \Big ( (\alpha-\alpha_{IR})^2 \Big ) \ . 
\label{beta_taylor_near_alfir}
\eeq
Here we have written this expansion for the full $\beta$ function; a
corresponding equation applies for $\beta_{n\ell}$.  

% ======================================================================

\subsection{IR Zero of $\beta$ at the Two-Loop Level}

We next review some background on the two-loop $\beta$ function that is
relevant for present work. The two-loop $\beta$ function is 
$\beta_{2\ell} = -2\alpha^2(\bar b_1 + \bar b_2 \alpha)$. 
This has an IR zero at 
\beq
\alpha_{IR,2\ell} = -\frac{\bar b_1}{\bar b_2} = -\frac{4\pi b_1}{b_2} \ , 
\label{alfir_2loop}
\eeq
which is physical if and only if $b_2 < 0$.  The coefficient $b_2$ is a linear,
monotonically decreasing function of $N_f$, which is positive for zero and
small $N_f$ and passes through zero, reversing sign, as $N_f$ increases through 
$N_{f,b2z}$, where 
\beq
N_{f,b2z} = \frac{17 C_A^2}{2T_f(5C_A+3C_f)} \ .
\label{nfb2z}
\eeq
For arbitrary $G$ and $R$, $N_{f,b1z} > N_{f,b2z}$, as is proved by the fact
that
\beq
N_{f,b1z} - N_{f,b2z} = \frac{3C_A(7C_A+11C_f)}{4T_f(5C_A+3C_f)} > 0 \ . 
\eeq
Hence, there is always an interval $I$ of $N_f$ values for which the 
two-loop $\beta$ function has an IR zero, namely
\beq
I: \quad N_{f,b2z} < N_f < N_{f,b1z} \ . 
\label{Nfinterval}
\eeq
For example, in the case of fermions in the fundamental representation, denoted
$\fund$, 
\beq
I: \quad \frac{34N_c^3}{13N_c^2-3} < N_f < \frac{11N_c}{2} \quad {\rm if} \ 
R = \fund \ ,
\label{Nfinterval_fund}
\eeq
so that, for $N_c=2$, the interval $I$ is $5.55 < N_f < 11$; for $N_c=3$, $I$
is $8.05 < N_f < 16.5$; and as $N_c \to \infty$, $I$ approaches $34N_c/13 < N_f
< 11N_c/2$.

Since we are primarily interested in studying the IR zero of $\beta$ and since
the presence or absence of an IR zero of the two-loop $\beta$ function,
$\beta_{2\ell}$, is a scheme-independent property, we focus on $N_f \in I$,
where this IR zero of $\beta_{2\ell}$ is present.  A general result is that for
a given gauge group $G$ and fermion representation $R$ and $N_f \in I$,
$\alpha_{IR,2\ell}$ is a monotonically decreasing function of $N_f$.  As $N_f$
decreases from $N_{f,b1z}$, $\alpha_{IR,2\ell}$ increases from 0. As $N_f$
decreases through a value labeled $N_{f,cr}$, $\alpha_{IR}$ increases through a
critical value, $\alpha_{cr} \sim O(1)$, where fermion condensation takes
place. Thus,
\beq
N_f = N_{f,cr} \ \Longleftrightarrow \ \alpha_{IR} = \alpha_{cr} \ . 
\label{nfdef}
\eeq
The value of $N_{f,cr}$ is of fundamental importance in the study of a
non-Abelian gauge theory, since it separates two different regimes of IR
behavior, viz., an IR conformal phase with no S$\chi$SB for $N_{f,cr} < N_f$ 
and an IR phase with S$\chi$SB for $N_f < N_{f,cr}$.  As $N_f$
approaches $N_{f,b2z}$ at the lower end of the interval $I$,
$\alpha_{IR,2\ell}$ becomes too large for Eq. (\ref{alfir_2loop}) to be
reliable.

Because of the strong-coupling nature of the physics at an approximate IR fixed
point with $\alpha_{IR} \sim O(1)$, there are significant higher-order
corrections to results obtained from the two-loop $\beta$ function, which
motivated the calculation of the location of the IR zero in $\beta$, and the
resultant value of $\gamma_m$ evaluated at this IR zero, to higher-loop order
for a general $G$, $R$, and $N_f$ \cite{bvh,ps}.

% ===========================================================================

\subsection{$\beta$ Function and Dilaton Mass} 

Here we focus on a theory in which the IR zero of $\beta$, $\alpha_{IR}$, is
slightly greater than $\alpha_{cr}$, so that, in the UV to IR flow, there is an
extended interval in $t=\ln \mu$ over which $\alpha(\mu)$ is approaching
$\alpha_{IR}$ from below, but is still less than $\alpha_{cr}$.  In this
interval, $\alpha(\mu) \sim O(1)$, but $\beta$ is small, and hence the theory
is approximately scale-invariant. As $\mu$ decreases through $\Lambda$, 
$\alpha(\mu)$ increases through $\alpha(\Lambda) = \alpha_{cr}$, the
fermion condensate forms, and the fermions gain dynamical masses, this
approximate scale invariance is broken spontaneously.  In terms of the
(symmetric) energy-momentum tensor $\theta^{\mu\nu}$, the dilatation current is
$D^\mu = \theta^{\mu\nu}x_\nu$, and one has $\partial_\mu D^\mu =
[\beta/(4\alpha)] G^a_{\mu\nu} G^{a \mu\nu}$, where $G^a_{\mu\nu}$ is the
field-strength tensor for the gauge field.  When taking matrix elements, the
deviation of this divergence $\partial_\mu D^\mu$ from zero, i.e., the
nonconservation of the dilatation current, thus arises from two sources, namely
the facts that $\beta$ is not exactly equal to zero and the nonzero value of
the matrix element of $G^a_{\mu\nu} G^{a\mu\nu}$, defined appropriately at
the scale $\Lambda$. An analysis of the matrix element of ${\cal D}^\mu$
between the vacuum and the dilaton state $|\chi(p)\rangle$, in conjunction with
a dimensional estimate of the gluon matrix element, and the Taylor series
expansion (\ref{beta_taylor_near_alfir}) evaluated with $\mu \sim \Lambda$
yields the resulting estimate for the dilaton mass $m_\chi$ \cite{dil}
\beq
m_\chi^2 \simeq \beta'_{IR} \, (\alpha_{IR}-\alpha_{cr}) \Lambda^2  \ . 
\label{mchisq}
\eeq
In terms of $n$-loop level quantities, the right-hand side of
Eq. (\ref{mchisq}) $\beta'_{IR,n\ell} \, (\alpha_{IR,n\ell}-\alpha_{cr})
\Lambda^2$. The importance of the slope at $\alpha_{IR}$, $\beta'_{IR}$, and
the $n$-loop calculation of this slope, $\beta'_{IR,n\ell}$, in estimating a
dilaton mass in a quasiconformal theory is evident from Eq. (\ref{mchisq}).  As
is the case with $\alpha_{IR,n\ell}$, because of the strong-coupling nature of
the physics, it is valuable to compute higher-loop corrections to the two-loop
result, $\beta'_{IR,2\ell}$.  Below, we will present two- and higher-loop
analytic and numerical calculations of $\beta'_{IR,n\ell}$.  Other effects on
$m_\chi$ have been discussed in the literature \cite{dil}, including the effect
of dynamical fermion mass generation associated with the spontaneous chiral
symmetry breaking as $\mu$ descends through the value $\Lambda$.  Owing to this
and other nonperturbative effects on $m_\chi$, we restrict ourselves here to
presenting one input to this calculation, namely $\beta'_{IR,n\ell}$, for which
we can give definite analytic and numerical results.

% =======================================================================

\subsection{IR Zero of $\beta$ at the Three-Loop Level} 

Let us assume that $N_f \in I$, so that $\beta_{2\ell}$ has an IR zero.  Here
we analyze how this IR changes as one calculates the $\beta$ function to
three-loop order, extending our results in \cite{bvh} to (an infinite set of)
schemes more general than the $\overline{MS}$ scheme used in that paper.  Since
the existence of of the IR zero in the two-loop $\beta$ function is a
scheme-independent property of the theory, it is reasonable to restrict to
schemes that preserve this IR zero of $\beta$ at the three-loop level.  We
first determine a condition for this to hold. 

The three-loop $\beta$ function is $\beta_{3\ell} = -2\alpha^2\beta_{3\ell,r}$,
so, aside from the double zero at $\alpha=0$ (the UV fixed point),
$\beta_{3\ell}$ vanishes at the two roots of the factor $\beta_{3\ell,r} \equiv
\bar b_1 + \bar b_2 \alpha + \bar b_3 \alpha^2 = 0$, namely,
\beqs
\alpha_{\beta z,3\ell,\pm} & = &
\frac{1}{2\bar b_3}\Big ( -\bar b_2 \pm
\sqrt{\Delta_2(\bar b_1,\bar b_2,\bar b_3)} \ \ \Big ) \cr\cr
& = &
\frac{2\pi}{b_3} \Big ( -b_2 \pm \sqrt{\Delta_2(b_1,b_2,b_3)} \ \ \Big ) \ .
\label{alf_betazero_3loop}
\eeqs
where $\Delta_2(b_1,b_2,b_3)=b_2^2-4b_1b_3$.  The analysis of the IR zero of
$\beta_{3\ell}$ requires an consideration of the sign of
$\Delta_2(b_1,b_2,b_3)$.  The condition that $\beta_{3\ell}$ have an IR zero
requires, in particular, that its two zeros away from the origin be real, i.e.,
that $\Delta_2(b_1,b_2,b_3) \ge 0$.  For a given $G$, $R$, and $N_f \in I$, so
that $b_1$ and $b_2$ are fixed, this condition amounts to an upper bound on
$b_3$, namely $b_3 \le b_2^2/(4b_1)$.  Now, $b_2 \to 0$ at the lower end of the
interval $I$, so that, insofar as one considers the analytic continuation of
$N_f$ from positive integers to positive real numbers, the above bound
generically requires that $b_3 \le 0$ for $N_f \in I$.  This is also required
if $R=\fund$, and one studies the theory in the
limit $N_c \to \infty$ and $N_f \to \infty$ with $r \equiv N_f/N_c$ fixed,
since in this case there are discrete pairs of values $(N_c,N_f)$ that enable
one to approach arbitrarily close to the lower end of the interval $I$ at
$r=34/13$ where $b_2 \to 0$. In order to preserve the existence of the two-loop
IR zero at the three-loop level, one is thus motivated to restrict to schemes
in which $b_3 \le 0$ for $N_f \in I$, and we will do so here. (The marginal
case $b_3=0$ is not generic, since $b_3$ varies as a function of $N_c$ and
$N_f$, so we will not consider it further.)

Before proceeding, it is worthwhile to recall how the property $b_3 < 0$ for
$N_f \in I$ arises in the $\overline{MS}$ scheme.  In this scheme, $b_3$ is a
quadratic function of $N_f$ with positive coefficients of the $N_f^2$ term and
the term independent of $N_f$.  This coefficient $b_3$ vanishes, with sign
reversal, at two values of $N_f$, denoted $N_{f,b3z,-}$ and $N_{f,b3z,+}$,
given as Eq. (3.16) in \cite{bvh}, with $b_3 < 0$ for $N_{f,b3z,-} < N_f <
N_{f,b3z,+}$ (and $b_3 > 0$ for $N_f < N_{f,b3z,-}$ and $N_f > N_{f,b3z,+}$.
In \cite{bvh} it was shown that in this scheme, for all of the representations
considered there, namely, the fundamental ($\fund$), adjoint, and rank-2
symmetric ($\sym$) and antisymmetric ($\asym$) tensor representations,
$N_{f,b3z,-} < N_{f,b2z}$ and $N_{f,b3z,2} > N_{f,b1z}$, so that $b_3 < 0$ for
all $N_f \in I$.  For example, for fermions in the $R=\fund$ representation,
(i) for $N_c=2$, $N_{f,b3z,1}=3.99 < N_{f,b2z}=5.55$ and $N_{f,b3z,2}=27.6 >
N_{f,b1z}=11$; (ii) for $N_c=3$, $N_{f,b3z,1}=5.84 < N_{f,b2z}=8.05$, and
$N_{f,b3z,2}=40.6 > N_{f,b1z}=16.5$; (iii) as $N_c \to \infty$, $N_{f,b3z,1}
\to 1.911N_c$ while $N_{f,b2z} \to 2.615N_c$ and $N_{f,b3z,2} \to 13.348N_c$,
while $N_{f,b1z} \to 5.5N_c$.  In Table \ref{discriminants} we list the values
of $\Delta_2(\bar b_1,\bar b_2,\bar b_3)$ with $\bar b_3$ calculated in the
$\overline{MS}$ scheme, for the illustrative cases $N_c=2, \ 3, \ 4$ and $N_f$
in the respective $I$ intervals.  Since $b_3 < 0$ for $N_f \in I$ in this
$\overline{MS}$ scheme, it follows that all of the entries in this table have
$\Delta_2 > 0$.

Given that $b_3 < 0$ for $N_f \in I$, Eq. (\ref{alf_betazero_3loop}) can be
rewritten as $\alpha = (2\pi/|b_3|)( -|b_2| \mp \sqrt{b_2^2 + 4b_1|b_3|})$.
The solution with a $-$ sign in front of the square root is negative and hence
unphysical; the other is positive and is $\alpha_{IR,3\ell}$, i.e.,
\beq
\alpha_{IR,3\ell} = \frac{2\pi}{|b_3|}
\Big ( -|b_2| + \sqrt{b_2^2 + 4b_1|b_3|} \ \ \Big ) \ . 
\label{alfir_3loop}
\eeq

In \cite{bvh} it was shown that in the $\overline{MS}$ scheme, for all $N_f \in
I$, $\alpha_{IR,3\ell} < \alpha_{IR,2\ell}$.  Here we demonstrate that this
result holds more generally than just in the $\overline{MS}$ scheme.  We prove
that for arbitrary gauge group $G$, fermion representation $R$, and $N_f \in
I$, in any scheme in which $b_3 < 0$ for $N_f \in I$ (which is thus guaranteed
to preserve the IR zero present at the two-loop level), it follows that
$\alpha_{IR,3\ell} < \alpha_{IR,2\ell}$. To prove this, we consider the
difference
\beq
\alpha_{IR,2\ell} - \alpha_{IR,3\ell} = \frac{2\pi}{|b_2 b_3|}
\bigg [ 2b_1 |b_3| + b_2^2 -|b_2|\sqrt{b_2^2+4b_1|b_3|} \ \bigg ] \ . 
\label{alfir_2loop_minus_alfir_3loop}
\eeq
The expression in square brackets is positive if and only if
\beq
(2b_1 |b_3|+b_2^2)^2-b_2^2(b_2^2+4b_1|b_3|) > 0 \ . 
\eeq
This difference is equal to the nonnegative quantity $(2b_1b_3)^2$, which
proves the inequality.  Note that, since $b_1$ is nonzero for asymptotic
freedom, this difference vanishes if and only if $b_3=0$, in which case
$\alpha_{IR,3\ell}=\alpha_{IR,2\ell}$. 
We have therefore proved that 
\beq
\alpha_{IR,3\ell} < \alpha_{IR,2\ell} \quad {\rm if} \ \ b_3 < 0 \quad {\rm
  for} \quad N_f \in I . 
\label{alfir_32loop_inequality}
\eeq

As noted above, $\alpha_{IR,2\ell}$ is a monotonically decreasing function of
$N_f \in I$.  With $b_3 < 0$ for $N_f \in I$, this monotonicity property is
also true of $\alpha_{IR,3\ell}$. As $N_f$ increases from $N_{f,b2z}$ to
$N_{f,b1z}$ in the interval $I$, $\alpha_{IR,3\ell}$ decreases from 
\beq
\alpha_{IR,3\ell} = 4\pi \sqrt{ \frac{b_1}{|b_3|}} \quad {\rm at} \ N_f =
N_{f,b2z} 
\label{alf_3loop_at_nfb2z}
\eeq
to zero as $N_f \nearrow N_{f,b1z}$ at the upper end of this interval,
vanishing like 
\beq
\alpha_{IR,3\ell} = \frac{4\pi b_1}{|b_2|} \bigg [1 - \frac{|b_3|b_1}{|b_2|^2} 
+ O( b_1^2 ) \ \bigg ]
\label{alfir_3loop_upperend}
\eeq
as $N_f \nearrow N_{f,b1z}$ and $b_1 \to 0$. 

% =======================================================================

\subsection{IR Zero of $\beta$ at the Four-Loop Level}

The four-loop $\beta$ function is $\beta_{4\ell} = -2\alpha^2 \beta_{4\ell,r}$,
so $\beta_{4\ell}$ has three
zeros away from the origin, at the roots of the cubic equation
\beq
\beta_{4\ell,r} \equiv \bar b_1 + \bar b_2\alpha + \bar b_3\alpha^2 + 
\bar b_4\alpha^3 = 0 \ , 
\label{eqbetazero_4loop_reduced}
\eeq
(where $\beta_{n\ell,r}$ was given in Eq. (\ref{beta_nloop_factor})). 
These zeros were analyzed for the $\overline{MS}$ scheme in \cite{bvh,ps}.
Here we extend this analysis to a more general class of
schemes that have $b_3 < 0$ for $N_f \in I$, and hence maintain at the
three-loop level the IR zero of the scheme-independent two-loop $\beta$
function.

The nature of the roots of Eq. (\ref{eqbetazero_4loop_reduced}) 
is determined by the sign of the discriminant 
$\Delta_3(\bar b_1,\bar b_2,\bar b_3,\bar b_4)$, or equivalently,
\beqs
\Delta_3 \equiv \Delta_3(b_1,b_2,b_3,b_4) & = & b_2^2b_3^2-27b_1^2b_4^2 - 
4(b_1b_3^3+b_4b_2^3) \cr\cr
& + & 18b_1b_2b_3b_4 \ . 
\label{disc3}
\eeqs
The following properties of $\Delta_3$ are relevant here: (i) if $\Delta_3 >
0$, then all of the roots of Eq. (\ref{eqbetazero_4loop_reduced}) are real;
(ii) if $\Delta_3 < 0$, then Eq.  (\ref{eqbetazero_4loop_reduced}) has one real
root and a complex-conjugate pair of roots; (iii) if $\Delta_3=0$, then at
least two of the roots of Eq. (\ref{eqbetazero_4loop_reduced}) coincide.  Given
the scheme-independent properties $b_1 > 0$ and $b_2 < 0$ (i.e., $N_f \in I$),
and provided that $b_3 < 0$ for $N_f \in I$, we can write this discriminant as
\beqs
\Delta_3(b_1,b_2,b_3,b_4) & = & b_2^2b_3^2-27b_1^2b_4^2 + 
4(b_1|b_3|^3+b_4|b_2|^3) \cr\cr
& + & 18b_1|b_2||b_3|b_4 \ . 
\label{disc3mag}
\eeqs
If $b_4$ were zero, then the zeros of $\beta_{4\ell}$ would coincide with those
of $\beta_{3\ell}$, and the property that these are all real is in accord
with the reduction 
\beq
\Delta_3(b_1,b_2,b_3,0) = b_3^2(b_2^2+4b_1|b_3|) = b_3^2\Delta_2(b_1,b_2,b_3) \
, 
\label{disc3_ifb4zero}
\eeq
which is positive.  

Now consider nonzero $b_4$.  First, assume that the scheme
has the property that $b_4 > 0$.  Then 
we can write Eq. (\ref{eqbetazero_4loop_reduced}) as 
\beq
\bar b_1-|\bar b_2|\alpha-|\bar b_3|\alpha^2+\bar b_4\alpha^3=0 \quad 
{\rm if} \ \ b_4 > 0 \ . 
\label{eqbetazero_4loop_reducedmag1}
\eeq
From an application of the Descartes theorem on roots of algebraic equations,
it follows that there are at most two (real) positive roots of this equation
and at most one negative root. Moreover, from Eq. (\ref{bnpos}) we can deduce
that in this case with $b_4 > 0$, in addition to the double zero at $\alpha=0$,
$\beta_{4\ell}$ has a zero at a negative value of $\alpha$, so the upper bound
on negative zeros from the Descartes theorem is saturated.  Furthermore, since
$\beta_{4\ell}$ is negative at large positive $\alpha$, there are then two
possibilities: either the two remaining zeros of Eq.
(\ref{eqbetazero_4loop_reduced}) are a complex-conjugate pair, or else they are
both real and positive.

If, on the other hand, the scheme is such that $b_4 < 0$, then we can write 
Eq. (\ref{eqbetazero_4loop_reduced}) as 
\beq
\bar b_1-|\bar b_2|\alpha-|\bar b_3|\alpha^2-|\bar b_4|\alpha^3=0 \quad 
{\rm if} \ \ b_4 < 0 \ . 
\label{eqbetazero_4loop_reducedmag2}
\eeq
From a similar application of the Descartes theorem, we infer that there is at
most one positive real root and at most two negative real roots of Eq.
(\ref{eqbetazero_4loop_reducedmag2}). From Eq. (\ref{bnneg}) we deduce that
$\beta_{4\ell}$ has a zero at a positive real value of $\alpha$.  Depending on
$|b_4|$, the other two roots of Eq. (\ref{eqbetazero_4loop_reducedmag2}) may be
real and negative or may form a complex-conjugate pair. 

Combining the information from both the Descartes theorem and the discriminant
$\Delta_3$, we derive the following conclusions about the roots of
Eq. (\ref{eqbetazero_4loop_reduced}) and hence the zeros of $\beta_{4\ell}$
aside from the double zero at $\alpha=0$.  As before, we assume that $b_2 < 0$
(i.e., $N_f \in I$) so that $\beta_{2\ell}$ has an IR zero, and also that the
scheme is such that $b_3 < 0$ for $N_f \in I$, guaranteeing that this IR zero
is maintained at the three-loop level.  Then,
\begin{enumerate} 

\item If $b_4 > 0$ and $\Delta_3 > 0$, then Eq. 
(\ref{eqbetazero_4loop_reduced}) has one negative and two positive real roots,

\item 

If $b_4 > 0$ and $\Delta_3 < 0$, then Eq. 
(\ref{eqbetazero_4loop_reduced}) has one negative root and a 
complex-conjugate pair of roots, 

\item 

If $b_4 < 0$ and $\Delta_3 > 0$, then Eq. 
(\ref{eqbetazero_4loop_reduced}) has one positive root and two negative 
roots, 

\item 

If $b_4 < 0$ and $\Delta_3 < 0$, then Eq. 
(\ref{eqbetazero_4loop_reduced}) has one positive root and a complex-conjugate
pair of roots. 

\end{enumerate}
For a particular pair $(N_c,N_f)$, the marginal case $\Delta_3=0$ might occur,
and would mean that two of the roots of Eq. (\ref{eqbetazero_4loop_reduced})
are degenerate. Since this equation is a cubic, it would follow that all of the
roots are real.  If $\Delta_3=0$ and $b_4 > 0$, then Eq.
(\ref{eqbetazero_4loop_reduced}) has one negative root and a positive root with
multiplicity 2, while if $b_4 < 0$, then (\ref{eqbetazero_4loop_reduced}) has
one positive root and a negative root with multiplicity 2. 

It is reasonable to avoid schemes that lead to the outcome (2) above, with no
real positive root of Eq. (\ref{eqbetazero_4loop_reduced}), since these fail to
preserve the IR zero of the scheme-independent two-loop $\beta$ function.
Although the positivity of $\Delta_3$ is not a necessary condition for this
preserving of the IR zero, it is a sufficient condition. We thus investigate
the conditions under which $\Delta_3$ is positive.  As shown via
Eq. (\ref{disc3_ifb4zero}), if $b_4=0$, then $\Delta_3 > 0$.  By continuity,
for small $|b_4|$, $\Delta_3$ remains positive, and there is only a small
shift in the two zeros that were present in $\beta_{3\ell}$, together with the
appearance of a new zero.  Since the highest-degree term in
$\Delta_3(b_1,b_2,b_3,b_4)$ involving $|b_4|$, namely $-27b_1^2b_4^2$ is
negative-definite, it follows that, other things being equal, for sufficiently
large $|b_4|$, $\Delta_3(b_1,b_2,b_3,b_4)$ will decrease through zero and
become negative.  The two $b_4$ values at which $\Delta_3(b_1,b_2,b_3,b_4)=0$
are
\beq
(b_4)_{\Delta 3z,\pm} = 
\frac{|b_2|(2b_2^2+9b_1|b_3|) \pm 2(b_2^2+3b_1|b_3|)^{3/2}}{27b_1^2}
\label{b3delta3zero}
\eeq
Therefore, a sufficient condition for a scheme to be such that $\beta_{4\ell}$
preserves the IR zero that is present in $\beta_{2\ell}$ and $\beta_{3\ell}$
for $N_f \in I$ is 
\beq
(b_4)_{\Delta 3z,-} < b_4 < (b_4)_{\Delta3z,+} \ . 
\label{b4_sufficient_range}
\eeq
Note that at the lower end of the interval $I$, where $b_2 \to 0$, the interval
(\ref{b4_sufficient_range}) reduces to the upper bound
$|b_4| < 2|b_3|^{3/2}/(27b_1)^{1/2}$. 

For reference, in the $\overline{MS}$ scheme, $b_4$ is a cubic polynomial in
$N_f$ and is positive for $N_f \in I$ for $N_c=2,3$ but is negative in part of
$I$ for higher values of $N_c$ (see Table I of \cite{bvh}, where $N_c$ is
denoted $N$). In Table \ref{discriminants} we list the values of $\Delta_3(\bar
b_1,\bar b_2,\bar b_3,\bar b_4)$ with $b_3$ and $b_4$ calculated in the
$\overline{MS}$ scheme, for the illustrative values $N_c=2, \ 3, \ 4$ and
values of $N_f$ in the respective $I$ intervals.  For all of the four-loop
entries in Table \ref{alfcritical_nloop}, $\Delta_3(\bar b_1,\bar b_2,\bar
b_3,\bar b_4) > 0$, as is evident from the values listed explicitly in Table
\ref{discriminants}, so these entries correspond to the case (1) in the list of
possibilities for $b_4$ and $\Delta_3$ given above.

Rather than calculating $\alpha_{IR,n\ell}$ directly from $\beta_{n\ell}$, a
different approach is to use $\beta_{n\ell}$ to compute Pad\'e approximants and
then calculate zeros of these approximants. As before, since one is interested
in the zeros away from the origin, one extracts the factor $-2\alpha^2$ in
Eq. (\ref{betanell}) and analyzes the polynomial $\beta_{n\ell,r}$ in
Eq. (\ref{beta_nloop_factor}), of degree $n-1$ in $\alpha$, depending on the
$n$ coefficients $\bar b_\ell$, $\ell=1,..,n$.  From this, one can construct a
set of $[p,q]$ Pad\'e approximants, i.e., rational functions, each with a
numerator polynomial of degree $p$ and a denominator polynomial of degree $q$
in $\alpha$, of the form $(\sum_{j=0}^p p_j z^j)/(\sum_{k=0}^q q_k z^k)$.
Without loss of generality, one can divide numerator and denominator by $q_0$,
so that, after redefinition of the coefficients, the $[p,q]$ Pad\'e approximant
to Eq. (\ref{beta_nloop_factor}) is
\beq
[p,q] = \frac{\sum_{j=0}^p p_j \alpha^j}
                  {1+\sum_{k=1}^q q_k \alpha^k} \ .
\label{pqpade}
\eeq
depending on the $p+q+1$ coefficients $p_j$ with $j=0,...,p$ and $q_k$ with
$k=1,..,q$.  These coefficients are determined by matching the Taylor series
expansion of $[p,q]$ in $\alpha$ with the $n$ coefficients $\bar b_\ell$,
$\ell=1,...,n$, so that $p+q=n-1$.  Thus, from the four-loop beta function
factor $\beta_{4\ell,r}$, one can construct two relevant approximants with
$p+q=3$, namely the [1,2] and [2,1] Pad\'e approximants \cite{otherpade}.  We
did this in \cite{bvh} and calculated the resultant unique IR zero from the
[1,2] approximant and the relevant IR zero from [2,1], denoted
$\alpha_{IR,4\ell,[1,2]}$ and $\alpha_{IR,4\ell,[2,1]}$, respectively.  These
were found to be close to the directly calculated IR zero, $\alpha_{IR,4\ell}$.
From the known results for $\beta_{n\ell}$ with $n=2, \ 3, \ 4$, one can make
estimates of $\beta_{5\ell}$ by various methods, but since these only contain
exact information up to the $n=4$ loop level, we will not pursue this direction
here.

As the $n=4$ special case of the result discussed above, 
$\alpha_{IR,4\ell}$ decreases to zero as $N_f \nearrow N_{f,b1z}$. 
For the $R=\fund$ and for a given $N_f \in I$ where
it is reliably calculable, $\alpha_{IR,4\ell}$ is slightly larger than
$\alpha_{IR,3\ell}$, but the difference, $\alpha_{IR,4\ell}-\alpha_{IR,3\ell}$
is sufficiently small that $\alpha_{IR,4\ell}$ is smaller than
$\alpha_{IR,2\ell}$.  For higher fermion representations and $N_f$ values where
the IR zero is reliably calculable (i.e., not too close to the lower end of the
interval $I$), the difference $\alpha_{IR,3\ell}-\alpha_{IR,4\ell}$ is again
smaller in magnitude then the difference $\alpha_{IR,2\ell}-\alpha_{IR,3\ell}$
but may have either sign. Thus, where $\alpha_{IR,4\ell}$ is reliably
calculable, it is smaller than $\alpha_{2,\ell}$.  The finding that the
fractional change in the location of the IR zero of $\beta$ is reduced at
higher-loop order agrees with the general expectation that calculating a
quantity to higher order in perturbation theory should give a more stable and
accurate result.  

The scheme-dependence of $\alpha_{IR,n\ell}$ for $n \ge 3$ can be studied by
carrying out scheme transformations, recalculating $\alpha'_{IR,n\ell}$ in the
new scheme, and comparing with $\alpha_{IR,n\ell}$. This study was carried out
in \cite{sch}.  To be acceptable, a scheme transformation must satisfy a number
of necessary conditions, such as mapping a a positive real $\alpha$ to a
positive real $\alpha'$ and vice versa.  Although these conditions can be
satisfied easily in the vicinity of the ultraviolet fixed point of an
asymptotically free theory at $\alpha=0$, they are nontrivial and constitute
significant restrictions on scheme transformations at an infrared fixed point
\cite{sch}. For example, the scheme transformation $\alpha = \tanh \alpha'$,
with inverse $\alpha' = (1/2)\ln[(1+\alpha)/(1-\alpha)]$, is acceptable for
small $\alpha$, in the vicinity of the UV fixed point of an asymptotically free
gauge theory, but is not acceptable in the vicinity of an IR fixed point at
$\alpha = \alpha_{IR} \sim O(1)$, since $\alpha$ can approach 1 from below, in
which case $\alpha'$ diverges, and $\alpha$ can exceed 1, in which case
$\alpha'$ is complex.

Scheme-dependence of higher-loop calculations is present not just in
calculations of an IR zero of $\beta_{n\ell}$ at three- and higher-loop level,
but also in higher-loop perturbative QCD calculations.  The fact that the
$\overline{MS}$ scheme is a reasonable one has been demonstrated, e.g., by the
excellent fit that has been obtained to experimental data for $\alpha_s(\mu)$
with $\mu^2=Q^2$ using this scheme \cite{bethke}. There has been much work on
optimized schemes for higher-order QCD calculations \cite{brodskyschemes}.
However, we note that two of the simplest scheme transformations that one might
apply for QCD are not generally acceptable at an IR zero of $\beta$ with
$\alpha_{IR} \sim O(1)$.  These are the scheme transformations denoted $S_2$
and $S_3$ in \cite{sch}, which are constructed to render the leading
scheme-dependent coefficient in the new scheme, $b_3'$, equal to zero. They are
acceptable at the UV zero of $\beta$ and hence in perturbative QCD
applications, but are not, in general, acceptable in the vicinity of an IR zero
with $\alpha_{IR} \sim O(1)$ because they can map a real positive $\alpha$ in
the $\overline{MS}$ scheme to a negative or complex coupling in the transformed
scheme, as was shown in \cite{sch}.

% =========================================================================

\subsection{Shift of IR Zero at $(n+1)$-loop Level}

Here we derive a result on the direction of the shift in the IR zero of the
$\beta$ function when one increases the order of calculation of $\beta$ from
the $n$-loop level to the $(n+1)$-loop level, where $n \ge 2$.  We assume, as
before, that the theory is asymptotically free and that $b_2 < 0$ (i.e., $N_f
\in I$), so that there is an IR zero of $\beta$ at the two-loop level.  We
assume that the scheme-dependent coefficients $b_\ell$ with $\ell=3,...,n+1$
are such that they preserve the existence of the IR zero of $\beta$ at
higher-loop level \cite{alt}.  We focus here on values of 
$\alpha$ close to $\alpha_{IR,n\ell}$, where $d\beta_{n\ell}/d\alpha > 0$.
Expanding $\beta_{n\ell}$ in a Taylor series expansion around
$\alpha=\alpha_{IR,n\ell}$, with the abbreviation 
$\beta'_{IR,n\ell} \equiv d\beta_{n\ell}/d\alpha {}|_{\alpha_{IR,n\ell}}$ 
defined above, we write the general Eq. 
(\ref{beta_taylor_near_alfir}) explicitly in terms of $n$-loop quantities as 
\beq
\beta_{n\ell} = \beta'_{IR,n\ell} \, (\alpha-\alpha_{IR,n\ell}) 
+ O \Big ( (\alpha-\alpha_{IR,n\ell})^2 \Big ) \ . 
\label{beta_taylor_near_alfir_nloop}
\eeq
Now let us calculate $\beta$ to the next-higher-loop order, i.e.,
$\beta_{(n+1)\ell}$, and solve for the zero, $\alpha_{IR,(n+1)\ell}$, which
corresponds to $\alpha_{IR,n\ell}$ (among the $n-1$ zeros of
$\beta_{(n+1)\ell}$ away from the origin). To determine whether
$\alpha_{IR,(n+1)\ell}$ is larger or smaller than $\alpha_{IR,n\ell}$, i.e.,
whether there is a shift to the right or left, consider the difference
\beq
\beta_{(n+1)\ell}-\beta_{n\ell} = -2 \bar b_{n+1} \alpha^{n+2} \ .
\label{beta_nplus1loop_minus_beta_nloop}
\eeq
In a scheme in which $b_{n+1} > 0$, this difference, evaluated at
$\alpha=\alpha_{IR,n\ell}$, is negative, so, given that
$d\beta_{n\ell}/d\alpha {}|_{\alpha_{IR,n\ell}} > 0$, to compensate 
for this, the zero shifts to the right, whereas if $b_{n+1} < 0$, the 
difference is positive, so the zero shifts to the left.  That is,
\beqs
{\rm If} \ b_{n+1} & > & 0 \ , \quad {\rm then} \ \alpha_{IR,(n+1)\ell} > 
\alpha_{IR,n\ell} \cr\cr
{\rm If} \ b_{n+1} & < & 0 \ , \quad {\rm then} \ \alpha_{IR,(n+1)\ell} <
\alpha_{IR,n\ell} \ . 
\label{alfir_nnp1_inequality}
\eeqs
(In a scheme with $b_{n+1}=0$, obviously
$\alpha_{IR,(n+1)\ell}=\alpha_{IR,n\ell}$).  The application of this general
result (\ref{alfir_nnp1_inequality}) is evident in the specific calculations in
\cite{bvh} at the three-loop and four-loop levels.

% ==========================================================================

\section{$\beta$ Function Structure}

At high scales in the UV, the $\beta$ function is dominated by the leading
quadratic term, $\beta \simeq - 2\bar b_1 \alpha^2 + O(\alpha^3)$.  The
calculation of the IR zero of $\beta_{n\ell}$ is important for investigating
the UV to IR evolution of the theory.  But, as discussed in the introduction,
for a more detailed study of this evolution, one needs not just the value of
the IR zero, $\alpha_{IR,n\ell}$, but the full curve of $\beta_{n\ell}$ for
$\alpha \in I_\alpha$. Here we present calculations of three quantities that
give further information about this curve, including (i) the value of $\alpha$
where $\beta_{n\ell}$ reaches its minimum for $\alpha \in I_\alpha$,
$\alpha_{m,n\ell}$, (ii) the minimum value of $\beta_{n\ell}$ for $\alpha \in
I_\alpha$, $(\beta_{n\ell})_{min}$; and the slope $\beta'_{IR,n\ell}$ at the IR
zero of $\beta$, as defined in Eq. (\ref{beta'_nloop}). The relevance of the
third quantity to estimates of the dilaton mass in a quasiconformal gauge
theory has been noted above. Our calculations are performed at the 
$n=2$, $n=3$, and $(n=4)$-loop level. 

% ==================================================================

\subsection{Two-Loop Level}

\subsubsection{Position of Minimum in $\beta_{2\ell}$ for $\alpha \in 
I_\alpha$} 

At the two-loop level, given that $b_2 < 0$ so that $\beta_{2\ell}$
function has an IR zero, the derivative 
$d\beta_{2\ell}/d\alpha = -2\alpha(2\bar b_1+3\bar b_2\alpha)$
vanishes at $\alpha=\alpha_{m,2\ell}$, where 
\beqs
\alpha_{m,2\ell} = -\frac{2\bar b_1}{3\bar b_2} = 
-\frac{8\pi b_1}{3b_2} = \frac{8\pi b_1}{3|b_2|} \ .
\label{alfcritical_2loop}
\eeqs
Explicitly, 
\beq
\alpha_{m,2\ell} = \frac{8\pi(11C_A-4T_fN_f)}
{3[4(5C_A+3C_f)T_fN_f-34C_A^2]} \ . 
\label{alf_critical_2loop_explicit}
\eeq
%

% ================================================================

\subsubsection{Minimum Value of $\beta_{2\ell}$ for $\alpha \in I_\alpha$ } 

At $\alpha=\alpha_{m,2\ell}$, $\beta_{2\ell}$ reaches its minimum physical
value for $\alpha \in I_\alpha$, namely 
\beqs
& & (\beta_{2\ell})_{min} = - \frac{8 \bar b_1^3}{27 \bar b_2^2} = 
- \frac{32 \pi b_1^3}{27 b_2^2} \cr\cr
& = & - \frac{32 \pi(11C_A-4T_fN_f)^3}
{81[34C_A^2-4(5C_A+3C_f)T_fN_f]^2} \ .
\label{beta_2loop_min}
\eeqs
Note that 
\beq
\alpha_{m,2\ell} = \frac{2}{3}\alpha_{IR,2\ell} \ .
\label{alfcritical_2loop_over_alfir_2loop}
\eeq
%

% ===================================================================

\subsubsection{Slope of $\beta_{2\ell}$ at $\alpha_{IR,2\ell}$}

The derivative $d\beta_{2\ell}/d\alpha$ evaluated at 
$\alpha=\alpha_{IR,2\ell}$ is 
\beqs
& & \beta'_{IR,2\ell} 
=-\frac{2\bar b_1^2}{\bar b_2}= -\frac{2b_1^2}{b_2} = \frac{2b_1^2}{|b_2|} 
 \cr\cr
& = & \frac{2(11C_A-4T_fN_f)^2}{3[4(5C_A+3C_f)T_fN_f-34C_A^2]} \ , 
\label{dbeta_2loop_dalf_at_alfir_2loop}
\eeqs
which is positive for $N_f \in I$.

As descriptors of the shape and structure of the $\beta$ function, the
quantities $\alpha_{m,n\ell}$, $(\beta_{n\ell})_{min}$, and $\beta'_{IR,n\ell}$
are interrelated.  Thus, if one makes a rough, linear ($lin.$) approximation
to the $\beta$ function in the interval from $\alpha=\alpha_{m,n\ell}$ to
$\alpha=\alpha_{IR,n\ell}$, then this slope would be
\beq
\frac{\Delta \beta_{n\ell;lin.}}{\Delta \alpha} =
\frac{-(\beta_{n\ell})_{min}}{\alpha_{IR,n\ell}-\alpha_{m,n\ell}} \ . 
\label{slope_linearapprox}
\eeq
For example, in the $(n=2)$-loop case, substituting the values of 
$(\beta_{2\ell})_{min}$, $\alpha_{IR,2\ell}$, and $\alpha_{m,2\ell}$, 
this approximation yields
\beq
\frac{\Delta \beta_{2\ell;lin.}}{\Delta \alpha} = -\frac{8\bar b_1^2}
{9\bar b_2} = \frac{8 b_1^2}{9|b_2|} \ ,
\label{slope_linearapprox_2loop}
\eeq
which exhibits the same dependence on the input coefficients $b_1$ and $b_2$,
with a somewhat smaller coefficient, $8/9$ rather than the coefficient 2 in 
the exact two-loop expression, $\beta'_{IR,2\ell}$, in Eq. 
(\ref{dbeta_2loop_dalf_at_alfir_2loop}). 

\begin{figure}
  \begin{center}
    \includegraphics[height=8cm,width=6cm]{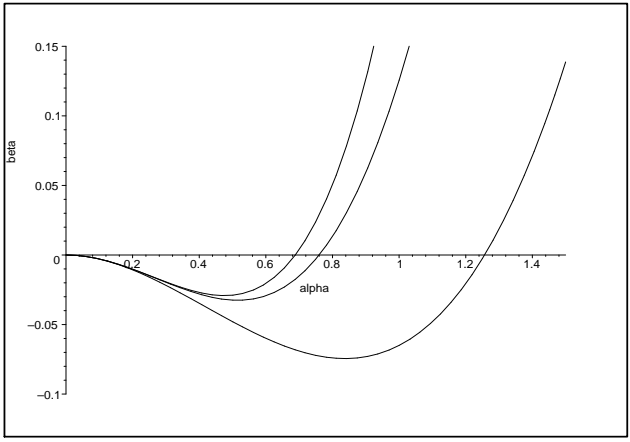}
  \end{center}
\caption{\footnotesize{ Plot of the $n$-loop $\beta$ function $\beta_{n\ell}$
as a function of $\alpha$ for $n=2, \ 3, \ 4$ and $N_c=2$, $N_f=8$ with
fermions in the fundamental representation. At a given value of
$\alpha$, the curves, from bottom to top, are for $\beta_{2\ell}$,
$\beta_{4\ell}$, and $\beta_{3\ell}$, respectively. See text for further
details.}}
\label{beta_Nc2Nf8}
\end{figure}

\begin{figure}
  \begin{center}
    \includegraphics[height=8cm,width=6cm]{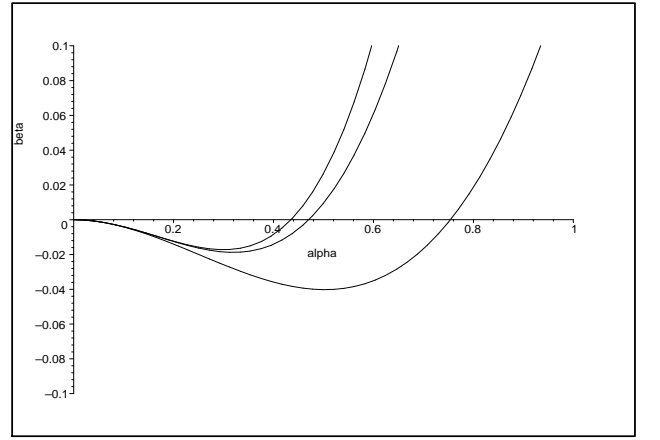}
  \end{center}
\caption{\footnotesize{
Plot of the $n$-loop $\beta$ function $\beta_{n\ell}$ as a function of
$\alpha$ for $n=2, \ 3, \ 4$ and the illustrative case $N_c=3$, $N_f=12$ with
fermions in the fundamental representation. At a given value of $\alpha$, 
the curves, from bottom to top, are for $\beta_{2\ell}$, $\beta_{4\ell}$, and 
$\beta_{3\ell}$, respectively. See text for further details.}}
\label{beta_Nc3Nf12}
\end{figure}

In Tables \ref{alfcritical_nloop}, \ref{beta_nloop_min}, and
\ref{dbeta_nloop_dalf_at_alfir_nloop} we list numerical values of
$\alpha_{m,n\ell}$, $(\beta_{n\ell})_{min}$, and $\beta'_{IR,n\ell}$ for
fermions in the $R=\fund$ representation of ${\rm SU}(N_c)$, for some
illustrative cases of $N_c$ and, for each $N_c$, values of $N_f$ in the
respective intervals $I$.  As illustrations, we show plots of $\beta_{n\ell}$
in Fig. \ref{beta_Nc2Nf8} for $N_c=2$ and $N_f=8$ and in
Fig. \ref{beta_Nc3Nf12} for $N_c=3$ and $N_f=12$ as functions of $\alpha$. The
results in the tables and figures are given for the quantities evaluated at the
$n=2$, $n=3$, and $n=4$ loop levels.  The $n=3$ and $n=4$ loop results will be
discussed further below.

%

% =============================================================================

\subsection{Three-Loop Level}

\subsubsection{Position of Minimum in $\beta_{3\ell}$ for $\alpha \in 
I_\alpha$} 

Here we assume a scheme in which $b_3 \ne 0$, since if
one is working with a scheme in which $b_3=0$, then $\beta_{3\ell} =
\beta_{2\ell}$, so the analysis of the three-loop $\beta$ function reduces to
that of the two-loop $\beta$ function discussed above. Furthermore, for the
reasons explained above, we restrict to schemes in which $b_3 < 0$ for $N_f \in
I$. The derivative $d\beta_{3\ell}/d\alpha =
-2\alpha(2\bar b_1+3\bar b_2\alpha+4\bar b_3\alpha^2)$ is zero at $\alpha=0$ 
and at the two other points,
\beq
\alpha = \frac{1}{8\bar b_3}\Big [ -3\bar b_2 \pm 
\sqrt{9\bar b_2^2-32\bar b_1\bar b_3} \ \ \Big ] \ . 
\label{a_dbeta_3loop_da_zero_gen}
\eeq
This can be written
as $\alpha = \pi(2|b_3|)^{-1}(-3|b_2| \mp \sqrt{9b_2^2+32b_1|b_3|})$. 
The critical point corresponding to the $-$ sign in front of the square root is
negative and hence unphysical, while the critical point corresponding to the
$+$ sign in front of the square root is $\alpha_{m,3\ell}$, i.e., 
\beq
\alpha_{m,3\ell} = \frac{\pi}{2|b_3|}\Big [ 
-3|b_2| + \sqrt{9b_2^2+32b_1|b_3|} \ \Big ] \ . 
\label{alf_critical_3loop}
\eeq

A general inequality is 
\beq
\alpha_{m,3\ell} < \alpha_{m,2\ell} \ . 
\label{alf_critical_23loop_dif_inequality}
\eeq
We prove this by examining the difference
\beqs
\alpha_{m,2\ell}-\alpha_{m,3\ell} & = & \frac{\pi}{6|b_2b_3|}\Big [ 
16b_1|b_3|+9b_2^2 \cr\cr
& - & 3|b_2|\sqrt{9b_2^2+32b_1|b_3|} \ \Big ]
\label{alf_critical2minus3}
\eeqs
The condition that this is positive is equivalent to the condition that the
square of the polynomial term in the numerator of
Eq. (\ref{alf_critical2minus3}) minus the square of the term in this numerator
involving the square root is positive.  This difference of squares is equal to
$256b_1^2b_3^2$, which is positive.  This proves the inequality
(\ref{alf_critical_23loop_dif_inequality}).

% ================================================================

\subsubsection{Minimum Value of $\beta_{3\ell}$ for $\alpha \in I_\alpha$ } 

At $\alpha=\alpha_{\beta_{min},3\ell}$, 
$\beta_{3\ell}$ reaches its minimum value for $\alpha \in I_\alpha$, namely
\beqs
(\beta_{3\ell})_{min} & = & \frac{\pi}{64|b_3|^3} \bigg [ 
-\Big ( 144b_1b_2^2|b_3| +128b_1^2b_3^2+27b_2^4 \Big ) \cr\cr
& + & |b_2|\Big (9b_2^2+32b_1|b_3| \Big )^{3/2} \ \bigg ] \ . 
\label{beta_3loop_min}
\eeqs
Note that one can write $(\beta_{3\ell})_{min}$ in terms of the $\bar b_\ell$
coefficients by replacing each $b_\ell$ in
Eq. (\ref{dbeta_3loop_dalf_at_alfir_3loop}) by the corresponding $\bar b_\ell$
and dividing the overall expression by $4\pi$.  Since $(\beta_{n\ell})_{min} <
0$, it is convenient to deal with the magnitudes $|(\beta_{n\ell})_{min}|$. We
find the following general inequality: for a given $G$, $R$, and $N_f \in I$,
in a scheme that has $b_3 < 0$ and hence maintains the existence of the IR zero
in $\beta_{2\ell}$,
\beq
|(\beta_{3\ell})_{min}| < |(\beta_{2\ell})_{min}| \ . 
\label{beta_23loop_min_inequality}
\eeq
To prove this, we consider the difference
\begin{widetext}
\beq
|(\beta_{2\ell})_{min}|-|(\beta_{3\ell})_{min}| = \frac{\pi}{1728b_2^2|b_3|^3}
 \bigg [ 2048b_1^3|b_3|^3+27b_2^2(144b_1b_2^2|b_3|+128b_1^2b_3^2+27b_2^4) -
27|b_2|^3(9b_2^2+32b_1|b_3|)^{3/2} \bigg ] \ . 
\label{beta_23loop_min_dif}
\eeq
\end{widetext}
The positivity of this difference is equivalent to the positivity of the square
of the polynomial terms in the numerator minus the square of the term in the
numerator involving the radical.  This difference of squares is equal to 
\beq
8192b_1^3|b_3|^3 \Big ( 512b_1^3|b_3|^3+3402 b_2^4b_1|b_3|+1728b_2^2b_1^2b_3^2
+729b_2^6) \ . 
\label{betamin23_difsquareterm}
\eeq
This expression is manifestly positive-definite, which proves the inequality
(\ref{beta_23loop_min_inequality}). 

% ==============================================================

\subsubsection{Slope of $\beta_{3\ell}$ at $\alpha_{IR,3\ell}$}

The derivative of $\beta_{3\ell}$ at $\alpha=\alpha_{IR,3\ell}$ is 
\beqs
\beta'_{IR,3\ell} & = & 
= \frac{1}{|b_3|^2} \Big [ -4|b_2|(b_2^2+b_1|b_3|) \cr\cr
& + & ( b_2^2+2b_1|b_3|) \sqrt{b_2^2+4b_1|b_3|} \ \ \Big ] \ .
\label{dbeta_3loop_dalf_at_alfir_3loop}
\eeqs
That this is positive follows from the fact that the square of the term
involving the square root minus the square of $-(4b_1|b_2||b_3| +|b_2|^3)$ in
the brackets is the manifestly positive quantity
$4b_1^2b_3^2(b_2^2+4b_1|b_3|)$.  Owing to the homogeneity properties, to
express $\beta'_{IR,3\ell}$ in terms of the $\bar b_\ell$ coefficients, one
simply replaces each $b_\ell$ in Eq. (\ref{dbeta_3loop_dalf_at_alfir_3loop}) by
the corresponding $\bar b_\ell$. 

A general inequality is that for a given $G$, $R$, and $N_f \in I$, in a scheme
with $b_3 < 0$, which is thus guaranteed to maintain the existence of the IR
zero in $\beta_{2\ell}$ at the three-loop level, 
\beq
\beta'_{IR,3\ell} < \beta'_{IR,2\ell} \ .
\label{dbeta_32loop_diff_inequality}
\eeq
To prove this, we examine the difference
\begin{widetext}
\beq
\beta'_{IR,2\ell} - \beta'_{IR,3\ell} = \frac{1}{|b_2|b_3^2} \Big [ 
2b_1^2b_3^2+4|b_2|(b_2^2+b_1|b_3|)-|b_2|(b_2^2+2b_1|b_3|)\sqrt{b_2^2+4b_1|b_3|}
\ \Big ] \ .
\label{betaprime23dif}
\eeq
\end{widetext}
The positivity of this difference is equivalent to the positivity of the square
of the polynomial terms in the numerator minus the square of the term in the
numerator involving the square root, which is 
\beq
4b_1^4 b_3^4+12 b_2^4 b_1^2 b_3^2+15 b_2^8+24 b_2^6 b1 |b_3| \ . 
\label{testprime}
\eeq
This is manifestly positive-definite, which proves the inequality
(\ref{dbeta_32loop_diff_inequality}).  

The shifts in the values of the IR zero,
$\alpha_{IR,n\ell}$, the position of the minimum in $\beta_{n\ell}$, the value
of $\beta_{n\ell}$ at the minimum, and the slope of $\beta_{n\ell}$ at
$\alpha=\alpha_{n,\ell}$ are evident from Tables \ref{alfcritical_nloop},
\ref{beta_nloop_min}, and \ref{dbeta_nloop_dalf_at_alfir_nloop} and
Figs. \ref{beta_Nc2Nf8} and \ref{beta_Nc3Nf12}.

% ===========================================================================

\subsection{Four-Loop Level}

The derivative $d\beta_{3\ell}/d\alpha = -2\alpha(2\bar b_1 + 3\bar
b_2\alpha+4\bar b_3\alpha^2+5\bar b_4\alpha^3)$ is zero at $\alpha=0$ and at
the three other points given by the zeros of the cubic equation $2\bar
b_1+3\bar b_2\alpha+4\bar b_3\alpha^2+5\bar b_4\alpha^3=0$.  We have calculated
these critical points, evaluated $\beta_{4\ell}$ at its minimum physical value,
and also evaluated the derivative $d\beta_{4\ell}/d\alpha$ at
$\alpha=\alpha_{IR,4\ell}$.  We give the numerical results for 
$\alpha_{m,4\ell}$, $(\beta_{4\ell})_{min}$, and $\beta'_{IR,4\ell}$ in 
Tables \ref{alfcritical_nloop}, \ref{beta_nloop_min}, and
\ref{dbeta_nloop_dalf_at_alfir_nloop}.  These four-loop structural results are
also evident in Figs. \ref{beta_Nc2Nf8} and \ref{beta_Nc3Nf12}. 

In addition to the results that we have proved above, we note some others here.
As stated above, numerical results for three- and four-loop structural
quantities were calculated in the $\overline{MS}$ scheme. First, although the
ratios $\alpha_{m,3\ell}/\alpha_{IR,3\ell}$ and
$\alpha_{m,4\ell}/\alpha_{IR,4\ell}$ are not constants as functions of $N_f$,
they do not differ very much from the two-loop ratio, which is a constant,
namely, 2/3, as given in Eq. (\ref{alfcritical_2loop_over_alfir_2loop}). 
With fermions in the $\fund$ representation, for a given $N_c$ and $N_f \in
I$, $\alpha_{m,4\ell}$ is slightly larger than $\alpha_{m,3\ell}$, but still
substantially smaller than $\alpha_{m,2\ell}$, just as is true of the
corresonding $\alpha_{IR,n\ell}$ quantities.  Moreover, 
for a given $G$ and loop order $n$, $\alpha_{m,n\ell}$ is a 
monotonically decreasing function of $N_f \in I$ and vanishes as 
$N_f \nearrow N_{f,b1z}$ and $b_1 \to 0$. 

% ==============================================================

\section{Some Properties of $\gamma_m$}

The anomalous dimension $\gamma_m$ for the fermion bilinear $\bar\psi\psi$
describes the scaling properties of this operator and
can be expressed as a series in $a$ or equivalently, $\alpha$:
\beq
\gamma_m = \sum_{\ell=1}^\infty c_\ell \, a^\ell
   = \sum_{\ell=1}^\infty \bar c_\ell \, \alpha^\ell \ , 
\label{gamma}
\eeq
where $\bar c_\ell = c_\ell/(4\pi)^\ell$ is the $\ell$-loop series coefficient.
The coefficient $c_1$ is scheme-independent, while $c_\ell$ for $\ell \ge 2$
are scheme-dependent. The $c_\ell$ coefficients have been calculated up 
to four-loop order in the $\overline{MS}$ scheme \cite{gamma4}. We list 
$c_\ell$ for $\ell=1, \ 2, \ 3$ in Appendix A. 
We denote the $n$-loop expression for $\gamma_m$ as a series in $\alpha$,
evaluated at the $n$-loop IR zero of $\beta$, $\alpha=\alpha_{IR,n\ell}$, as
$\gamma_{IR,n\ell}$. 

In \cite{bvh}, we calculated $\gamma_{IR,n\ell}$ up to $(n=4)$-loop order in
the $\overline{MS}$ scheme. An important result was that we found a substantial
reduction in $\gamma_{IR}$ going from the two-loop to three-loop level for all
of the fermion representations that were considered.  The difference going from
three- to four-loop level, $\gamma_{IR,3\ell}-\gamma_{IR,4\ell}$, was found to
be smaller and could be of either sign, depending on the representation and
value of $N_f$.  The resultant $\gamma_{IR,4\ell}$ was thus substantially
smaller than $\gamma_{IR,2\ell}$.

One may investigate the reduction 
$\gamma_{IR,3\ell} < \gamma_{IR,2\ell}$ found in \cite{bvh} further.  To do 
this for a given gauge group $G$ and fermion representation $R$, we assume that
$N_f \in I$, so that the theory has an IR zero of $\beta_{2\ell}$, and,
further, that the scheme is such that $b_3 < 0$ for $N_f \in I$, so that this
IR zero is guaranteed to be maintained at the three-loop level.  We will use
the resultant property that $\alpha_{IR,3\ell} < \alpha_{IR,2\ell}$.  Let us
consider the difference $\gamma_{IR,2\ell} - \gamma_{IR,3\ell}$. This is given
by 
\beqs
& & \gamma_{IR,2\ell} - \gamma_{IR,3\ell} =
\bar c_1(\alpha_{IR,2\ell}-\alpha_{IR,3\ell}) \cr\cr
& + & 
\bar c_2(\alpha_{IR,2\ell}^2-\alpha_{IR,3\ell}^2) -
\bar c_3 \alpha_{IR,3\ell}^3 \ . 
\label{gamma23difexplicit}
\eeqs
The (scheme-independent) coefficient $c_1$ is positive, so that, since
$\alpha_{IR,2\ell}-\alpha_{IR,3\ell} > 0$, it follows that the first term on
the right-hand side of Eq. (\ref{gamma23difexplicit}) is positive.  The factor
$(\alpha_{IR,2\ell}^2-\alpha_{IR,3\ell}^2)$ in the second term is also
positive.  The coefficient $c_2$ is scheme-dependent, so the analysis
of this term necessarily involves a choice of scheme, as does the analysis of
the third term.  We next prove that in the $\overline{MS}$ scheme, 
$c_2 > 0$ for all of the representations considered in \cite{bvh}, so that this
second term is positive.  To show this, we begin with the $\fund$
representation, for which 
\beq
(c_2)_{fund,\overline{MS}} = \frac{(N_c^2-1)(203N_c^2-9-20N_cN_f)}{192N_c^2} 
\ . 
\label{c2_fund_msbar}
\eeq
The first factor in the numerator, $N_c^2-1$, is obviously positive for all 
physical $N_c$. The second factor is positive for $N_f < N_{f,c2z}$, where
\beq
N_{f,c2z} = \frac{203N_c^2-9}{20N_c} \ . 
\label{nfc2z}
\eeq
This is larger than the upper bound on $N_f$ from asymptotic freedom,
$N_{f,b1z}$, as is clear from the difference
\beq
N_{f,c2z} - N_{f,b1z} = \frac{3(31N_c^2-3)}{20N_c} > 0 \ ,
\label{nfc2zminusnfb1z}
\eeq
so that $c_{2,fund,{\overline{MS}}} > 0$ for $N_f \in I$ (actually for all
physical $N_f$).  This provides an analytic understanding of the numerical
results in Table V of \cite{bvh}, which indicated that $c_2 > 0$ for all $N_c$
and $N_f$ considered there.

We next consider the case of fermions in the adjoint representation, for which 
\beq 
c_{2,adj,\overline{MS}} = \frac{N_c^2(53-10N_f)}{24} \ . 
\label{c2_adj_msbar}
\eeq
This is positive for $N_f < 53/10$, which is larger than the upper bound on 
$N_f$ for this representation from the requirement of asymptotic freedom, 
namely $N_{f,b1z}=11/4= 2.75$, so again, $c_2 > 0$ for all $N_f$ for the
adjoint representation in the $\overline{MS}$ scheme. 

Finally, we consider the case of fermions in the symmetric or antisymmetric
rank-2 tensor representation, denoted $S2$ and $A2$, respectively, with Young
tableaux $\sym$ and $\asym$. Owing to the fact that the $A2$ representation of
SU(2) is the singlet, it is understood that $N_c \ge 3$ in this case.  Since
various formulas are similar for these two representations, with appropriate
reversals of signs of certain terms, it is convenient to give them in a unified
manner, with $T2$ referring to $S2$ and $A2$ together.  We have
\begin{widetext}
\beq 
c_{2,T2,\overline{MS}} = \frac{(N_c \pm 2)(N_c \mp 1)\Big [ 109N_c^2 \pm 9N_c -
    18 - 10N_c(N_c \pm 2)N_f]}{48N_c^2} \ , 
\label{c2_t2_msbar}
\eeq
\end{widetext}
where the upper (lower) sign applies for the $S2$ ($A2$) representation,
respectively. In the numerator of this expression, the factor 
$(N_c \pm 2)(N_c \mp 1)$ is obviously positive for the relevant values of
$N_c$, so one next examines the factor 
$[109N_c^2 \pm 9Nc - 18 - 10N_c(Nc \pm 2)N_f]$.  This is positive for 
$N_f < N_{f,c2T2z}$, where 
\beq
N_{f,c2T2z} = \frac{106N_c^2 \pm 9N_c - 18}{10N_c(N_c \pm 2)} \ . 
\eeq
As for the other representations, $N_{f,c2T2z}$ is larger than the respective 
upper bound on $N_f$ from the requirement of asymptotic freedom,
$N_{f,b1z,T2}$,
\beq
N_{f,b1z,T2} = \frac{11(N_c}{2(N_c \pm 2)} \ .
\label{nfb1zt2}
\eeq
This is proved by considering the difference 
\beq
N_{f,c2T2z}-N_{f,b1z,T2} = \frac{3(17N_c^2 \pm 3N_c - 6)}{10N_c(N_c\pm 2)} 
\label{nfc2t2z_minus_nfb1zt2}
\eeq
This difference is positive for 
\beq
N_c \ge \frac{\mp 3 + \sqrt{417}}{34} \ , 
\eeq
i.e., 0.5124 for $S2$ and 0.6888 for $A2$, and hence for all physical $N_c$.
Therefore, this proves that $c_2 > 0$ for all relevant $N_f < N_{f,b1z}$ and,
in particular, for all $N_f$ in the respective intervals $I$ for these theories
with fermions in the symmetric or antisymmetric rank-2 representation.

We have thus proved that for the $\overline{MS}$ scheme, for all of the
representations considered in \cite{bvh}, the first two terms in the difference
$\gamma_{IR,2\ell}-\gamma_{IR,3\ell}$ are both positive.  We have also
investigated the contribution of the third term.  By analytic methods similar
to those exhibited above, we find that this third term also makes a positive
contribution to the difference in Eq. (\ref{gamma23difexplicit}), i.e., $c_3 <
0$ for $N_f \in I$, in most, although not all, cases. For example, for $G={\rm
SU}(N_c)$ and fermions in the $\fund$ representation, $c_3 < 0$ for 
all $N_c$ up to $N_c=15$ and integer $N_f$ values in the respective
intervals $I$. This includes all of the cases of $N_c$ and $N_f \in I$ for
which numerical results were given in \cite{bvh} and thus gives an analytic
understanding of those results.  For $N_c=16$, the interval $I$ is $42 \le N_f
\le 87$, and $c_3 < 0$ for all of these values of $N_f$ except the lowest one,
$N_f=42$, where $c_3 > 0$.  Similar comments apply for larger $N_c$.

% ==================================================================

\section{Supersymmetric Gauge Theory}

\subsection{IR Zeros of $\beta$} 

It is of interest to give some corresponding results on properties of the
$\beta$ function and associated UV to IR evolution in an asymptotically free,
${\cal N}=1$ supersymmetric gauge theory with vectorial chiral superfield
content $\Phi_i, \ \tilde \Phi_i$, $i=1,...,N_f$ in the $R, \ \bar R$
representations, respectively.  A number of exact results have been derived
describing the infrared properties of the theory in
\cite{nsvzbeta,seiberg}. Thus, one can compare findings from perturbative
calculations with these exact results, and this was done in \cite{bfs}.  The
$\beta$ function of the theory has the form
\beq
\beta_s = \frac{d\alpha}{dt} = -2\alpha \sum_{\ell=1}^\infty b_{\ell,s} a^\ell 
= -2\alpha \sum_{\ell=1}^\infty \bar b_{\ell,s} \alpha^\ell \ , 
\label{beta_susy}
\eeq
where we use the subscript $s$, standing for supersymmetric, to avoid
confusion with the corresponding quantities in the nonsupersymmetric theory,
and $\bar b_{\ell,s} = b_{\ell,s}/(4\pi)^\ell$.  The beta function
calculated to $n$-loop order is denoted $\beta_{n\ell,s}$.  For values of $N_f$
for which $\beta_{n\ell,s}$ has an IR zero, we denote this as
$\alpha_{IR,n\ell,s}$.  In addition to the scheme-independent coefficients
$b_{1,s}$ and $b_{2,s}$, calculated in \cite{b1s} and \cite{b2s}, respectively,
the three-loop coefficient, $b_{3,\ell}$, has been calculated in \cite{b3s} in
the dimensional reduction ($\overline{DR}$) scheme \cite{dred}.  Calculations
of $\alpha_{IR,n\ell,s}$ and corresponding values of the anomalous dimension
for the bilinear chiral superfield operator $\Phi \tilde \Phi$ were given in
\cite{bfs} up to the maximal order to which $b_\ell$ and the coefficients of
the anomalous dimension had been calculated, namely the three-loop level.

We recall that, since $b_1=3C_A - 2T_fN_f$ \cite{b1s}, the upper bound on $N_f$
for the theory to be asymptotically free is
\beq
N_f < N_{f,b1z,s} \ , 
\label{nfupperbound}
\eeq
where
\beq
N_{f,b1z,s} = \frac{3C_A}{2T_f} \ . 
\label{nfb1z_susy}
\eeq
The two-loop $\beta$ function coefficient is $b_{2,s}=6C_A^2-4(C_A+2C_f)T_fN_f$
\cite{b2s}, which decreases through positive values and passes through zero,
reversing sign, as $N_f$ increases through
\beq
N_{f,b2z,s} = \frac{3C_A^2}{2T_f(C_A+2C_f)} \ . 
\label{nfb2z_susy}
\eeq
Since $N_{f,b2z,s} < N_{f,b1z,s}$, there is always an interval of values of
$N_f$, namely
\beq
I_s: \quad N_{f,b2z,s} < N_f < N_{f,b1z,s} \ ,
\label{Nfinterval_susy}
\eeq
in which the two-loop $\beta$ function for this
theory has an IR zero for $N_f$. The value of this two-loop IR zero is 
\beqs
\alpha_{IR,2\ell,s} & = & -\frac{4\pi b_{1,s}}{b_{2,s}} \cr\cr
       & = & \frac{2\pi(3C_A-2T_fN_f)}{2(C_A+2C_f)T_fN_f-3C_A^2} \ .
\label{alfir2loop_susy}
\eeqs
In particular, for chiral superfields in the $\fund$ and $\bar{\fund}$ 
representations, $N_{f,b1z,s}=3N_c$ and $N_{f,b2z,s} = 3N_c^3/[2N_c^2-1]$, 
so
\beq
I_s: \frac{3N_c^3}{2N_c^2-1} < N_f < \frac{3N_c}{2} 
\quad {\rm for} \ R = \fund \ . 
\label{Nfinterval_fund_susy}
\eeq
In this case, the exact value of $N_f$ at the lower end of the IR-conformal,
non-Abelian Coulomb phase was determined in Ref. \cite{seiberg} to be
\beq
N_{f,cr,s}= \frac{3N_c}{2} \quad {\rm for} \ R = \fund \ . 
\label{nfcr_fund_susy}
\eeq
Here, since $N_{f,b2z,s} > N_{f,cr,s}$, the coefficient $b_{2,s}$ passes
through zero and reverses sign in the interior of the non-Abelian Coulomb
phase.  Consequently, as was noted in \cite{bfs}, for this case of chiral
superfields in the $\fund$ and $\bar{\fund}$ representations, one cannot study
the IR zero of $\beta_{2\ell}$ throughout the entirety of this phase.  The
generalization of $N_{f,cr,s}$ to higher representations has been given as
\cite{rsbeta}
\beq
N_{f,cr,s} = \frac{3C_A}{2T_f} \ .
\label{nfcr_susy}
\eeq
Assuming that $\gamma_m$ saturates its upper bound of 1 in this supersymmetric
theory as $N_f \searrow N_{f,cr,s}$, Eq. (\ref{nfcr_fund_susy}) agrees with the
result obtained via a closed-form solution for $\beta$ \cite{nsvzbeta} 
(see also \cite{sth}).  
Note that the fact that the
scheme used in \cite{nsvzbeta} is different from the $\overline{DR}$ scheme
does not affect this, since $N_{f,cr,s}$ is a physical quantity.

At the three-loop level, $\beta_{3\ell,s}=-2\alpha^2 \beta_{3\ell,r,s}$, where
$\beta_{n\ell,r}$ is given by Eq. (\ref{beta_nloop_factor}) with $\bar b_\ell$
replaced by $\bar b_{\ell,s}$.  One makes use of the result \cite{b3s}
\beqs
b_{3,s} & = & 21C_A^3 + 4(-5C_A^2-13C_AC_f+4C_f^2)T_fN_f \cr\cr
    & + & 4(C_A+6C_f)T_f^2N_f^2 
\label{b3s}
\eeqs
in the $\overline{DR}$ scheme. The three-loop IR zero of $\beta_s$,
$\alpha_{IR,3\ell,s}$, was calculated and compared with $\alpha_{IR,2\ell,s}$
in \cite{bfs}. One can prove various inequalities similar to those that we have
proved above for a non-supersymmetric gauge theory.  We illustrate one of
these, concerning the relative size of $\alpha_{IR,2\ell,s}$ and
$\alpha_{IR,3\ell,s}$ for chiral superfields in the $\fund$ and $\bar{\fund}$
representations. We begin by noting that $b_{3,s}$ in the $\overline{DR}$
scheme is a quadratic function of $N_f$ which is positive for small $N_f$, and,
as $N_f$ increases, passes through zero, becoming negative, at a value denoted
$N_{f,b3z,-,s}$, reaches a minimum, and then passes through zero again at
$N_f=N_{f,b3z,+,s}$, and is positive for larger $N_f$.  In general,
\beq
N_{f,b3z,\pm,s} = \frac{5C_A^2+13C_AC_f-4C_f^2 \pm \sqrt{R_s} }
{2T_f(C_A+6C_f)} \ , 
\label{nfb3z_pm_susy}
\eeq
where
\beq
R_s = 4C_A^4 + 4C_A^3C_f + 129 C_A^2 C_f^2 - 104C_AC_f^3 + 16C_f^4 \ . 
\label{rs}
\eeq
For the $\fund$ or $\bar{\fund}$ representation, this reduces to 
\beq
N_{f,b3z,\pm,s} = \frac{21N_c^4-9N_c^2-2 \pm \sqrt{R_{s,fund}}}
{2N_c(4N_c^2-3)} \ , 
\label{nfb3z_pm_fund_susy} 
\eeq
where
\beq
R_{s,fund} = 105N_c^8-126N_c^6-3N_c^4+36N_c^2+4 \ . 
\label{rsfund}
\eeq
(Note that $R_{s,fund}$ is positive-definite, vanishing at eight complex
values of $N_f$.)  To prove that $\alpha_{IR,2\ell,s} < \alpha_{IR,3\ell,s}$
for this case, it suffices to show that $b_{3,s} < 0$ for $N_f \in I_s$, since
then one can apply the same proof that we used for
Eq. (\ref{alfir_32loop_inequality}). To show that $b_{3,s} < 0$ for $N_f \in
I_s$, we will demonstrate that $N_{f,b3z,-,s} < N_{f,b2z,s}$ and $N_{f,b3z,+,s}
> N_{f,b1z,s}$.  First, for this case of $R$ equal to the
$\fund$ representation, we consider the difference
\begin{widetext}
\beq
N_{f,b2z,s} - N_{f,b3z,-,s} =  \frac{-18N_c^6+21N_c^4-5N_c^2-1 +
  (2N_c^2-1)\sqrt{R_{s,fund}}}{2N_c(2N_c^2-1)(4N_c^2-3)} \ . 
\label{nfb2z_fund_susy_minus_nfb3zminus_susy}
\eeq
\end{widetext}
Although the polynomial term in the numerator of
(\ref{nfb2z_fund_susy_minus_nfb3zminus_susy}) is negative, it is smaller than
the term involving the square root.  To show this, we observe that the square
of the term involving the square root minus the square of the polynomial term
in the numerator is $24Nc^4(N_c^2+1)(N_c^2-1)^2(4N_c^2-3)$.  This is positive
for all physical $N_c$, proving that $N_{f,b2z,s} > N_{f,b3z,-,s}$ for this
case.  Next, we consider the difference
\beq
N_{f,b3z,+,s}-N_{f,b1z} = \frac{-3N_c^4+9N_c^2-2 + \sqrt{R_{s,fund}}}
{2N_c(4N_c^2-3)} \ . 
\label{nfb3zplus_susy_minus_nfb1z_susy}
\eeq
Although the polynomial term in the numerator is negative for physical $N_c$,
it is smaller than the square root, as is shown by the fact that the difference
of the square of the square root term minus the square of the polynomial term
is $24N_c^2(N_c^4-1)(4N_c^2-3)$, which is positive.  So we have proved that for
this case, $N_{f,b3z,-,s} < N_{f,b2z,s}$ and $N_{f,b3z,+,s} > N_{f,b1z,s}$. 
In turn, this proves that for this case with $N_f$ chiral
superfields in the $\fund$ and $\bar{\fund}$ representations, in the
$\overline{DR}$ scheme, $b_{3,s} < 0$ for $N_f \in I_s$, and hence
\beq
\alpha_{IR,3\ell,s} < \alpha_{IR,2\ell,s} \ .
\label{alfineq_susy}
\eeq
This inequality follows by the same type of proof as the one that we gave for 
Eq. (\ref{alfir_32loop_inequality}). 

\subsection{Structural Properties of $\beta_s$}

Because one has exact, nonperturbative results available for this theory, we
will be brief in our discussion of structural properties of the $\beta$
function.  The value of $\alpha$ where $\beta_{2\ell,s}$ has zero slope and a
minimum in the interval (\ref{alphainterval}) is given by
Eq. (\ref{alfcritical_2loop}) as
\beq
\alpha_{m,2\ell,s} = \frac{8\pi(3C_A-2T_fN_f)}
{3[4(C_A+2C_f)T_fN_f-6C_A^2]} \ . 
\label{alf_critical_2loop_explicit_susy}
\eeq
At this $\alpha$, $\beta_{2\ell,s}$ reaches its minimum physical value,
\beq
(\beta_{2\ell,s})_{min} = - \frac{32\pi(3C_A-2T_fN_f)^3}
{27[4(C_A+2C_f)T_fN_f-6C_A^2]^2} \ .
\label{beta_2loop_min_susy}
\eeq
As in the nonsupersymmetric theory, $\alpha_{m,2\ell,s} = 
(2/3)\alpha_{IR,2\ell,s}$. 

The derivative $d\beta_{2\ell,s}/d\alpha$ evaluated at
$\alpha=\alpha_{IR,2\ell,s}$ is given by the 
analogue of Eq. (\ref{dbeta_2loop_dalf_at_alfir_2loop}), namely 
\beq
\beta'_{IR,2\ell,s} = \frac{2(3C_A-2T_fN_f)^2}{4(C_A+2C_f)T_fN_f-6C_A^2} \ , 
\label{dbeta_2loop_dalf_at_alfir_2loop_susy}
\eeq
which is positive for $N_f \in I$.  

At the three-loop level, $\alpha_{m,3\ell,s}$, $(\beta_{3\ell})_{min}$, and
$\beta'_{IR,3\ell,s}$ are given by Eqs. (\ref{alf_critical_3loop}),
(\ref{beta_3loop_min}), and (\ref{dbeta_3loop_dalf_at_alfir_3loop}) with the
replacements $b_\ell \to b_{\ell,s}$.

\begin{figure}
  \begin{center}
    \includegraphics[height=8cm,width=6cm]{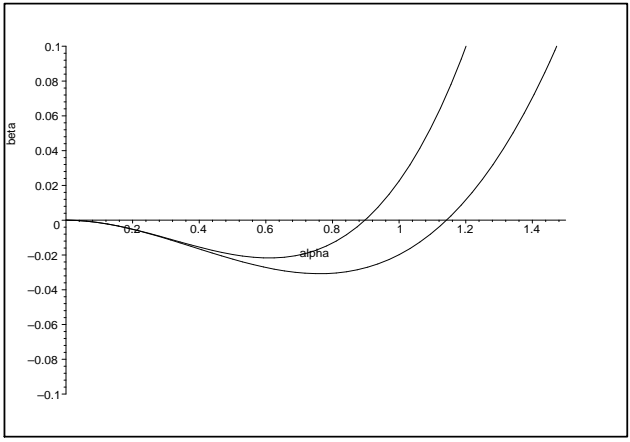}
  \end{center}
\caption{\footnotesize{ Plot of the $n$-loop $\beta$ function $\beta_{n\ell,s}$
for an SU($N_c$) gauge theory with ${\cal N}=1$ supersymmetry, as a function of
$\alpha$, for $n=2$ and $n=3$ loops and $N_c=2$, $N_f=5$ with chiral
superfields in the fundamental representation.  The lower and upper curves
correspond to $\beta_{2\ell,s}$ and $\beta_{3\ell,s}$, respectively.  See text
for further details.}}
\label{betasusy_Nc2Nf5}
\end{figure}

\begin{figure}
  \begin{center}
    \includegraphics[height=8cm,width=6cm]{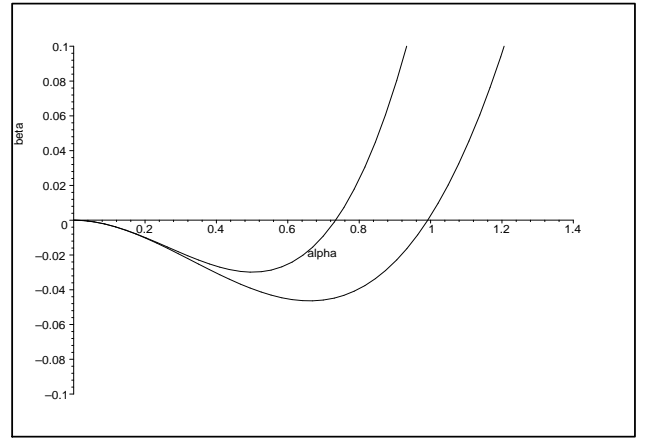}
  \end{center}
\caption{\footnotesize{ Plot of the $n$-loop $\beta$ function $\beta_{n\ell,s}$
for an SU($N_c$) gauge theory with ${\cal N}=1$ supersymmetry, as a function of
$\alpha$, for $n=2$ and $n=3$ loops and $N_c=3$, $N_f=7$ with chiral
superfields in the fundamental representation.  The lower and upper curves
correspond to $\beta_{2\ell,s}$ and $\beta_{3\ell,s}$, respectively.  See text
for further details.}}
\label{betasusy_Nc3Nf7}
\end{figure}

In Figs. \ref{betasusy_Nc2Nf5} and \ref{betasusy_Nc3Nf7} we show plots of the
two-loop and three-loop $\beta$ functions for this supersymmetric gauge theory
with chiral superfields in the fundamental representation and with the
illustrative values $N_c=2$, $N_f=5$ and $N_c=3$, $N_f=7$, respectively.  The
three-loop $\beta$ functions are calculated in the $\overline{DR}$ scheme.
%

% ===============================================================

\section{Conclusions}

In this paper we have studied some higher-loop structural properties of the
$\beta$ function in an asymptotically free vectorial gauge theory, focusing on
the case where the theory has an IR zero in the $\beta$ function. These
structural properties include the value of $\alpha$ where $\beta$ reaches a
minimum (i.e., a maximal magnitude, since $\beta \le 0$ for $\alpha \in
I_\alpha$), the value of $\beta$ at this minimum, and the derivative
$d\beta/d\alpha$ at the IR zero, calculated to the $n$-loop order.  We have
given results up to four loops in a non-supersymmetric gauge theory and
up to three loops in a gauge theory with ${\cal N}=1$ supersymmetry. In an
asymptotically free theory with an exact or approximate infrared zero in the
$\beta$ function, these structural quantities provide further information about
the running of $\alpha$ as a function of the reference scale, $\mu$. The
derivative of $\beta$ at $\alpha_{IR}$ is also of interest because it enters
into estimates of the dilaton mass in a quasiconformal gauge theory.  A general
inequality was proved concerning how the shift in the IR zero of $\beta$ as one
goes from the $n$-loop to the $(n+1)$-loop order depends on the sign of
$b_{n+1}$.  For schemes which have $b_3 < 0$ for $N_f \in I$ and which thus are
guaranteed to preserve the existence of the IR zero in the (scheme-independent)
$\beta_{2\ell}$ at the three-loop level, we have proved that $\alpha_{IR,3\ell}
< \alpha_{IR,2\ell}$, $\alpha_{m,3\ell} < \alpha_{m,2\ell}$,
$|(\beta_{3\ell})_{min}| < |(\beta_{2\ell})_{min}|$, and $\beta'_{IR,3\ell} <
\beta'_{IR,2\ell}$.  Our results further elucidate the ultraviolet to infrared
evolution of an asymptotically free vectorial gauge theory with fermions.

% ================================================================

\bigskip
\bigskip

Acknowledgments: I would like to thank T. Ryttov for collaboration on the
earlier works \cite{bvh,bfs,sch}, and T. Appelquist and the theory group at
Yale for useful discussions and warm hospitality during the sabbatical period
when this work was done.  This research was partially supported by the grant
NSF-PHY-09-69739.

\bigskip
\bigskip
\bigskip

% ================================================================

\section{Appendix A}
\label{betagamma}

\subsection{$\beta$ Function Coefficients} 

For a vectorial gauge theory with gauge group $G$ and $N_f$ fermions in the
representation $R$, the coefficients $b_1$ and $b_2$ in the $\beta$ function 
are \cite{b1}
\beq
b_1 = \frac{1}{3}(11 C_A - 4T_fN_f)
\label{b1}
\eeq
and \cite{b2,casimir}
\beq
b_2=\frac{1}{3}\left [ 34 C_A^2 - 4(5C_A+3C_f)T_fN_f \right ]
\ .
\label{b2}
\eeq
In the $\overline{MS}$ scheme \cite{b3}
\beqs
b_3 & = & \frac{2857}{54}C_A^3 +
+ T_fN_f \bigg[ 2C_f^2 - \frac{205}{9} C_AC_f - \frac{1415}{27}C_A^2 \bigg ] 
\cr\cr
& + & (T_fN_f)^2 \bigg [ \frac{44}{9}C_f + \frac{158}{27}C_A \bigg ] \ . 
\label{b3}
\eeqs
We have also used the four-loop coefficient, $b_4$, calculated in the 
$\overline{MS}$ scheme in \cite{b4}, for our calculations. This coefficient 
$b_4$ is a cubic polynomial in $N_f$.  

\subsection{Coefficients for $\gamma_m$ } 

We list here the $c_\ell$ for $\ell=1, \ 2, \ 3$: 
\beq
c_1 = 6C_f
\label{c1}
\eeq
\beq
c_2 = 2C_f \Big [ \frac{3}{2}C_f + \frac{97}{6}C_A  - \frac{10}{3}
 T_fN_f \Big ] \ .
\label{c2}
\eeq
\begin{widetext}
\beq
c_3 = 2C_f \bigg [ \frac{129}{2}C_f^2 - \frac{129}{4}C_fC_A +
\frac{11413}{108} C_A^2 + C_f(T_fN_f) \Big (-46+48\zeta(3) \Big )
-C_A(T_fN_f)\Big (\frac{556}{27}+48\zeta(3)\Big )-\frac{140}{27} (T_fN_f)^2 
\bigg ] \ , 
\label{c3}
\eeq
where $\zeta(s) = \sum_{n=1}^\infty n^{-s}$ is the Riemann zeta function, and
$\zeta(3)=1.2020569...$.  We have also used the four-loop coefficient 
$c_4$, calculated in the $\overline{MS}$ scheme in \cite{gamma4}, for our 
calculations. 

\end{widetext}

% =================================================================

\section{Appendix B}
\label{betafunctions}

Here we give some illustrative explicit $\beta_{4\ell}$ functions for 
various values of $N_c$ and $N_f$.  The three-loop and four-loop coefficients
are calculated in the $\overline{MS}$ scheme.  These are written in the form
\beq
\beta_{4\ell} = -2\bar b_1 \alpha^2 \Big [ 1 + \sum_{\ell=2}^4 \Big ( 
\frac{\bar b_\ell}{\bar b_1} \Big ) \alpha^{\ell-1} \Big ] 
\label{beta4ellform}
\eeq
and are listed both analytically and numerically (to the indicated
floating-point accuracy).

\begin{widetext}
\beqs
N_c=2, \ N_f=8: \quad \beta_{4\ell} & = & -\frac{\alpha^2}{\pi} \bigg [ \ 1 
- \frac{5}{2}    \Big (\frac{\alpha}{\pi} \Big ) 
- \frac{603}{64} \Big (\frac{\alpha}{\pi} \Big )^2
+ \Big (\frac{-136859+198528\zeta(3)}{9216}\Big )
\Big (\frac{\alpha}{\pi} \Big )^3 \ \bigg ] \cr\cr
& = & -0.3183\alpha^2\Big ( 1 - 0.7958\alpha - 0.9546\alpha^2 +
0.3562\alpha^3 \Big )
\label{beta_4loop_Nc2Nf8}
\eeqs
\beqs
N_c=2, \ N_f=9: \quad \beta_{4\ell} & = & -\frac{2\alpha^2}{3\pi} \bigg [ \ 1 
- \frac{169}{32}    \Big (\frac{\alpha}{\pi} \Big ) 
- \frac{154445}{9216} \Big (\frac{\alpha}{\pi} \Big )^2
+ \Big (\frac{-22506041+50531904\zeta(3)}{1327104}\Big )
\Big (\frac{\alpha}{\pi} \Big )^3 \ \bigg ] \cr\cr
& = & -0.2122\alpha^2\Big ( 1 - 1.6811\alpha - 1.6980\alpha^2 +
0.9292\alpha^3 \Big )
\label{beta_4loop_Nc2Nf9}
\eeqs

\beqs
N_c=3, \ N_f=10: \quad \beta_{4\ell} & = & -\frac{13\alpha^2}{6\pi} \bigg [ \ 1
- \frac{37}{26}    \Big (\frac{\alpha}{\pi} \Big )
- \frac{41351}{3744} \Big (\frac{\alpha}{\pi} \Big )^2
+ \Big (\frac{-13418011+13331592\zeta(3)}{404352}\Big )
\Big (\frac{\alpha}{\pi} \Big )^3 \ \bigg ] \cr\cr
& = & -0.6897\alpha^2\Big ( 1 - 0.4530\alpha - 1.1191\alpha^2 +
0.2080 \alpha^3 \Big )
\label{beta_4loop_Nc3Nf10}
\eeqs
\beqs
N_c=3, \ N_f=12: \quad \beta_{4\ell} & = & -\frac{3\alpha^2}{2\pi} \bigg [ \ 1
- \frac{25}{6}    \Big (\frac{\alpha}{\pi} \Big )
- \frac{6361}{288} \Big (\frac{\alpha}{\pi} \Big )^2
+ \Big (\frac{-140881+219192\zeta(3)}{3456}\Big )
\Big (\frac{\alpha}{\pi} \Big )^3 \ \bigg ] \cr\cr
& = & -0.4775\alpha^2\Big ( 1 - 1.3263\alpha - 2.2379\alpha^2 +
1.1441 \alpha^3 \Big ) \ . 
\label{beta_4loop_Nc3Nf12}
\eeqs
\end{widetext}
%

% =================================================================

\section{Appendix C}
\label{disc}

Consider the polynomial of degree $m$ in $z$, $P_m(z) = \sum_{s=0}^m \kappa_s
z^s$.  As discussed in Section \ref{betafunction}, information on the nature of
the roots of the equation $P_m(z)=0$ is given by the discriminant $\Delta_m$
defined in Eq. (\ref{deltam}).  Since $\Delta_m$ is a symmetric function of the
roots (being proportional to the square of the Vandermonde polynomial of these
roots), the theorem on symmetric functions \cite{symfun} implies that
$\Delta_m$ can be expressed as a polynomial in the coefficients $\kappa_s$,
$s=0,...,m$.  We indicate this in the notation $\Delta_m =
\Delta_m(\kappa_0,...,\kappa_m)$. The discriminant $\Delta_m$ is most
conveniently calculated in terms of the Sylvester matrix of $P(z)$ and
$dP(z)/dz$, equivalent to the resultant matrix, denoted $S_{P,P'}$, of
dimension $(2m-1) \times (2m-1)$:
\beq
\Delta_m = (-1)^{m(m-1)/2}\kappa_m^{-1}{\rm det}(S_{P,P'}) \ . 
\label{Deltammatrix}
\eeq
Since we will use $\Delta_m$ for $m=2$ and $m=3$, we
list the explicit expressions here: 
\beq
\Delta_2(\kappa_0,\kappa_1,\kappa_2) = \kappa_1^2-4\kappa_0\kappa_2 \ . 
\label{deltam2}
\eeq
For $m=3$,
\beq
S_{P_3,P_3'} = \left( \begin{array}{ccccc}
\kappa_3  & \kappa_2  & \kappa_1  & \kappa_0  &    0     \\
 0        & \kappa_3  & \kappa_2  & \kappa_1  & \kappa_0 \\
3\kappa_3 & 2\kappa_2 & \kappa_1  &    0      &    0     \\
  0       & 3\kappa_3 & 2\kappa_2 & \kappa_1  &  0       \\
  0       &   0       & 3\kappa_2 & 2\kappa_2 & \kappa_1 
\end{array} \right )
\label{sp3pp3}
\eeq
so that 
\beqs
\Delta_3(\kappa_0,\kappa_1,\kappa_2,\kappa_3) & = & 
(\kappa_1 \kappa_2)^2 - 27(\kappa_0 \kappa_3)^2 
- 4(\kappa_0 \kappa_2^3+\kappa_3 \kappa_1^3) \cr\cr
& + & 18 \kappa_0 \kappa_1 \kappa_2 \kappa_3 \ .
\label{deltam3}
\eeqs

\newpage

% table I: alfcritical_nloop and alfir_nloop
%
\begin{table}
\caption{\footnotesize{Value of $\alpha_{m,n\ell}$ at the $n$-loop level with 
$n=2, \ 3, \ 4$ for an SU($N_c$) gauge theory with $N_f$ fermions in the
fundamental representation, with $N_f \in I$. As discussed in the text,
$\alpha_{m,n\ell}$ is the value at which the $n$-loop $\beta$
function takes on its minimum value in the interval $0 \le \alpha \le
\alpha_{IR,n\ell}$.  Results are given for the illustrative values $N_c=2, \ 3,
\ 4$. For comparison, we also list the IR zeros of $\beta$ calculated at
$n$-loop level, $\alpha_{IR,n\ell}$, for $n=2, \ 3, \ 4$, from Ref. \cite{bvh}.
For this and other tables, quantities evaluated at the $n=3$ and $n=4$ loop
level are calculated in the $\overline{MS}$ scheme.}}
\begin{center}
\begin{tabular}{|c|c|c|c|c|c|c|c|} \hline\hline
$N_c$ & $N_f$ & 
$\alpha_{m,2\ell}$ & 
$\alpha_{m,3\ell}$ & 
$\alpha_{m,4\ell}$ & 
$\alpha_{IR,2\ell}$ & 
$\alpha_{IR,3\ell}$ & 
$\alpha_{IR,4\ell}$ 
\\ \hline
 2  &  7  &  1.89   & 0.735  & 0.823  & 2.83   & 1.05   & 1.21    \\
 2  &  8  &  0.838  & 0.476  & 0.515  & 1.26   & 0.688  & 0.760   \\
 2  &  9  &  0.397  & 0.286  & 0.300  & 0.595  & 0.418  & 0.444   \\
 2  & 10  &  0.154  & 0.133  & 0.135  & 0.231  & 0.196  & 0.200   \\
\hline
 3  & 10  &  1.47   & 0.534  & 0.563  & 2.21   & 0.764  & 0.815   \\
 3  & 11  &  0.823  & 0.402  & 0.429  & 1.23   & 0.579  & 0.626   \\
 3  & 12  &  0.503  & 0.300  & 0.320  & 0.754  & 0.435  & 0.470   \\
 3  & 13  &  0.312  & 0.217  & 0.228  & 0.468  & 0.317  & 0.337   \\
 3  & 14  &  0.185  & 0.146  & 0.151  & 0.278  & 0.215  & 0.224   \\
 3  & 15  &  0.0952 & 0.0834 & 0.0846 & 0.143  & 0.123  & 0.126   \\
 3  & 16  &  0.0277 & 0.0416 & 0.0267 & 0.0416 & 0.0397 & 0.0398  \\
\hline
 4  & 13  &  1.23   & 0.422  & 0.436  & 1.85   & 0.604  & 0.628   \\
 4  & 14  &  0.773  & 0.340  & 0.359  & 1.16   & 0.489  & 0.521   \\
 4  & 15  &  0.522  & 0.275  & 0.293  & 0.783  & 0.397  & 0.428   \\
 4  & 16  &  0.364  & 0.221  & 0.235  & 0.546  & 0.320  & 0.345   \\
 4  & 17  &  0.256  & 0.174  & 0.184  & 0.384  & 0.254  & 0.271   \\
 4  & 18  &  0.177  & 0.133  & 0.138  & 0.266  & 0.194  & 0.205   \\
 4  & 19  &  0.117  & 0.0954 & 0.0981 & 0.175  & 0.140  & 0.145   \\
 4  & 20  &  0.0697 & 0.0613 & 0.0621 & 0.105  & 0.0907 & 0.0924   \\
 4  & 21  &  0.0315 & 0.0472 & 0.0297 & 0.0472 & 0.044  & 0.0444  \\
\hline\hline
\end{tabular}
\end{center}
\label{alfcritical_nloop}
\end{table}
%

% table II: Delta values 
%
\begin{table}
\caption{\footnotesize{Values of the discriminants $\Delta_2(\bar b_1,\bar
b_2,\bar b_3)$ and $\Delta_3(\bar b_1,\bar b_2,\bar b_3,\bar b_4)$ 
(see Eqs. (\ref{deltam}) and (\ref{deltambbar})) for the
three-loop and four-loop IR zero equations, with $\bar b_3$ and $\bar b_4$
calculated in the $\overline{MS}$ scheme. Results are given for the
illustrative values $N_c=2, \ 3, \ 4$.
Notation $a$e-n means $a \times 10^{-n}$.}}
\begin{center}
\begin{tabular}{|c|c|c|c|} \hline\hline
$N_c$ & $N_f$ & $\Delta_2(\bar b_1,\bar b_2,\bar b_3)$ &  
$\Delta_3(\bar b_1,\bar b_2,\bar b_3,\bar b_4)$
\\ \hline
 2  &  7  &  0.107    & 0.151e-2   \\
 2  &  8  &  0.113    & 0.399e-2   \\
 2  &  9  &  0.108    & 0.885e-2   \\
 2  & 10  &  0.0963   & 1.68e-2    \\
\hline
 3  & 10  &  0.557    & 0.0943     \\
 3  & 11  &  0.596    & 0.170      \\
 3  & 12  &  0.610    & 0.293      \\
 3  & 13  &  0.603    & 0.493      \\
 3  & 14  &  0.577    & 0.803      \\
 3  & 15  &  0.537    & 1.221      \\
 3  & 16  &  0.489    & 1.676      \\
\hline
 4  & 13  &  1.75     & 1.53       \\
 4  & 14  &  1.87     & 2.45       \\
 4  & 15  &  1.95     & 3.74       \\
 4  & 16  &  1.97     & 5.60       \\
 4  & 17  &  1.96     & 8.32       \\
 4  & 18  &  1.92     & 12.18      \\
 4  & 19  &  1.85     & 17.36      \\
 4  & 20  &  1.75     & 23.71      \\
 4  & 21  &  1.64     & 30.59      \\
\hline\hline
\end{tabular}
\end{center}
\label{discriminants}
\end{table}
%

% table III: minimum value of beta on physical interval 
%
\begin{table}
\caption{\footnotesize{Minimum value of the $n$-loop $\beta$
    function, $\beta_{n\ell}$, denoted
$(\beta_{n\ell})_{min}$, in the interval
$0 \le \alpha \le \alpha_{IR,n\ell}$ relevant for the UV to IR evolution,
calculated to $n=2, \ 3, \ 4$ loop order for an SU($N_c$) theory with 
$N_f$ fermions in the fundamental representation, with $N_f \in I$. Values are
    given for $N_c=2, \ 3, \ 4$. Notation $a$e-n means $a \times 10^{-n}$.}}
\begin{center}
\begin{tabular}{|c|c|c|c|c|} \hline\hline
$N_c$ & $N_f$ & 
$(\beta_{2\ell})_{min}$ &  
$(\beta_{3\ell})_{min}$ & 
$(\beta_{4\ell})_{min}$ 
\\ \hline
 2  &  7  &  $-0.504$       & $-0.998$e-1   & $-0.117$      \\
 2  &  8  &  $-0.745$e-1    & $-0.292$e-1   & $-0.326$e-1   \\
 2  &  9  &  $-1.11$e-2     & $-0.660$e-2   & $-0.703$e-2   \\
 2  & 10  &  $-0.836$e-3    & $-0.666$e-3   & $-0.680$e-3   \\
\hline
 3  & 10  &  $-0.498$       & $-0.863$e-1   & $-0.934$e-1   \\
 3  & 11  &  $-1.32$e-1     & $-0.394$e-1   & $-0.432$e-1   \\
 3  & 12  &  $-0.4025$e-1   & $-1.72$e-2    & $-1.88$e-2    \\
 3  & 13  &  $-1.20$e-2     & $-0.672$e-2   & $-0.719$e-2   \\
 3  & 14  &  $-0.304$e-2    & $-0.209$e-2   & $-0.218$e-2   \\
 3  & 15  &  $-0.481$e-3    & $-0.392$e-3   & $-0.399$e-3   \\
 3  & 16  &  $-1.36$e-5     & $-1.28$e-5    & $-1.28$e-5    \\
\hline
 4  & 13  &  $-0.484$       & $-0.752$e-1   & $-0.790$e-1   \\
 4  & 14  &  $-0.169$       & $-0.419$e-1   & $-0.452$e-1   \\
 4  & 15  &  $-0.674$e-1    & $-0.232$e-1   & $-0.252$e-1   \\
 4  & 16  &  $-0.2815$e-1   & $-1.24$e-2    & $-1.35$e-2    \\
 4  & 17  &  $-1.16$e-2     & $-0.6215$e-2  & $-0.667$e-2   \\
 4  & 18  &  $-0.444$e-2    & $-0.280$e-2   & $-0.296$e-2   \\
 4  & 19  &  $-1.45$e-3     & $-1.055$e-3   & $-1.09$e-3    \\
 4  & 20  &  $-0.343$e-3    & $-0.282$e-3   & $-0.287$e-3   \\
 4  & 21  &  $-0.350$e-4    & $-0.319$e-4   & $-0.321$e-4   \\
\hline\hline
\end{tabular}
\end{center}
\label{beta_nloop_min}
\end{table}
%

% table IV: slope of the beta function 
%
\begin{table}
\caption{\footnotesize{Value of $d\beta_{n\ell}/d\alpha$ at 
$n=2, \ 3, \ 4$ loop order for an SU($N_c$) theory with $N_f$ fermions in the 
fundamental representation, with $N_f \in I$, evaluated at the IR zero
calculated to this order, $\alpha_{IR,n\ell}$. We denote this here as 
$\beta'_{IR,n\ell}$.}}
\begin{center}
\begin{tabular}{|c|c|c|c|c|} \hline\hline
$N_c$ & $N_f$ & 
$\beta'_{IR,2\ell}$ &  
$\beta'_{IR,3\ell}$ &  
$\beta'_{IR,4\ell}$ 
\\ \hline
 2  &  7  &  1.20   & 0.728  & 0.677   \\
 2  &  8  &  0.400  & 0.318  & 0.300   \\
 2  &  9  &  0.126  & 0.115  & 0.110   \\
 2  & 10  &  0.0245 & 0.0239 & 0.0235  \\
\hline
 3  & 10  &  1.52   & 0.872  & 0.853   \\
 3  & 11  &  0.720  & 0.517  & 0.498   \\
 3  & 12  &  0.360  & 0.2955 & 0.282   \\
 3  & 13  &  0.174  & 0.156  & 0.149   \\
 3  & 14  &  0.0737 & 0.0699 & 0.0678  \\
 3  & 15  &  0.0227 & 0.0223 & 0.0220  \\
 3  & 16  &  0.00221& 0.00220& 0.00220 \\
\hline
 4  & 13  &  1.77   & 0.965  & 0.955   \\
 4  & 14  &  0.984  & 0.655  & 0.639   \\
 4  & 15  &  0.581  & 0.440  & 0.424   \\
 4  & 16  &  0.348  & 0.288  & 0.276   \\
 4  & 17  &  0.204  & 0.180  & 0.1725  \\
 4  & 18  &  0.113  & 0.105  & 0.101   \\
 4  & 19  &  0.0558 & 0.0536 & 0.0522  \\
 4  & 20  &  0.0222 & 0.0218 & 0.0.215 \\
 4  & 21  &  0.00501& 0.00499& 0.00496 \\
\hline\hline
\end{tabular}
\end{center}
\label{dbeta_nloop_dalf_at_alfir_nloop}
\end{table}


\bigskip
\bigskip


\begin{thebibliography}{99}

\bibitem{b1}
D. J. Gross and F. Wilczek, Phys. Rev. Lett. {\bf 30}, 1343 (1973); 
H. D. Politzer, Phys. Rev. Lett. {\bf 30}, 1346 (1973); G. 't Hooft,
unpublished. 

\bibitem{b2}
W. E. Caswell, Phys. Rev. Lett. {\bf 33}, 244 (1974); 
D. R. T. Jones, Nucl. Phys. B {\bf 75}, 531 (1974). 

\bibitem{b3}
O. V. Tarasov, A. A. Vladimirov, and A. Yu. Zharkov, Phys. Lett. B {\bf 93},
429 (1980); S. A. Larin and J. A. M. Vermaseren, Phys. Lett. B {\bf 303}, 334
(1993). 

\bibitem{b4}
T. van Ritbergen, J. A. M. Vermaseren, and S. A. Larin, Phys. Lett. B {\bf
400}, 379 (1997). 

\bibitem{ms}
G. 't Hooft, Nucl. Phys. B {\bf 61}, 455 (1973).

\bibitem{msbar}
W. A. Bardeen, A. J. Buras, D. W. Duke, and T. Muta, Phys. Rev. D {\bf 18},
3998 (1978).

\bibitem{gamma4}
J. A. M. Vermaseren, S. A. Larin, and T. van Ritbergen, Phys. Lett. B {\bf
405}, 327 (1997).

\bibitem{mf}
%
It is straightforward to include fermion masses, since these are
gauge-invariant in a vectorial theory.  However, for a given fermion mass $m$,
as the reference scale $\mu$ decreases below $m$, one would integrate these out
of the low-energy effective theory applicable for $\mu < m$, so a massive
fermion would not affect the UV to IR evolution significantly below its mass. 

\bibitem{bz}
T. Banks and A. Zaks, Nucl. Phys. B {\bf 196}, 189 (1982). 

\bibitem{gk98}
E. Gardi and M. Karliner, Nucl. Phys. B {\bf 529}, 383 (1998);
E. Gardi, G. Grunberg and M. Karliner, JHEP {\bf 9807}, 007 (1998).

\bibitem{bvh}
T. A. Ryttov and R. Shrock, Phys. Rev. D {\bf 83}, 056011 (2011), 
arXiv:1011.4542.

\bibitem{ps}
C. Pica and R. Sannino, Phys. Rev. D {\bf 83}, 035013 (2011), 
arXiv:1011.5917. 

\bibitem{bfs}
T. A. Ryttov and R. Shrock, Phys. Rev. D {\bf 85}, 076009 (2012),
arXiv:1202.1297. 

\bibitem{sch}
T. A. Ryttov and R. Shrock, 
Phys. Rev. D {\bf 86}, 065032 (2012), arXiv:1206.2366; 
Phys. Rev. D {\bf 86}, 085005 (2012), arXiv:1206.6895. 

\bibitem{alfdif34}
%
For the three-loop versus four-loop comparison, the difference
$\alpha_{IR,3\ell}-\alpha_{IR,4\ell}$ was found in \cite{bvh} to be smaller in
magnitude than $\alpha_{IR,2\ell}-\alpha_{IR,3\ell}$, and was negative for the
fundamental representation but could be either positive or negative for a
higher representation such as the adjoint.

\bibitem{dilconf}
J.-F. Fortin, B. Grinstein, and A. Stergiou, arXiv:1208.3674. 

\bibitem{otherwtc}
K. Yamawaki, M. Bando, and K. Matumoto, Phys. Rev. Lett. {\bf
56}, 1335 (1986); B. Holdom, Phys. Rev. Lett. {\bf 60}, 1233 (1988);
V. Miransky and K. Yamawaki, Phys. Rev. D {\bf 55}, 5051
(1997). 

\bibitem{chipt}
T. Appelquist, D. Karabali, and L. C. R. Wijewardhana, Phys. Rev. Lett. {\bf
57}, 957 (1986); T. Appelquist and L. C. R. Wijewardhana, Phys. Rev. D
{\bf 35}, 774 (1987); Phys.  Rev. D {\bf 36}, 568 (1987);
T. Appelquist, J. Terning, and L. C. R. Wijewardhana,
Phys. Rev. Lett. {\bf 77}, 1214 (1996). 

\bibitem{sg}
T. Appelquist and F. Sannino, Phys. Rev. D {\bf 59}, 067702 (1999);
M. Harada, M. Kurachi, and K. Yamawaki, Phys. Rev. {\bf 70}, 033009 (2004);
M. Kurachi and R. Shrock, Phys. Rev. D {\bf 74}, 056003 (2006);
M. Kurachi and R. Shrock, JHEP {\bf 12}, 034 (2006);
M. Kurachi, R. Shrock, and K. Yamawaki, Phys. Rev. D {\bf 76}, 035003 (2007).

\bibitem{lgt}
%
T. Appelquist, G. Fleming, E. Neil, and D. Schaich, 
Phys. Rev. D {\bf 84}, 054501 (2011); 
T. DeGrand, Phys. Rev. D {\bf 84}, 116901 (2011); 
Z. Fodor, K. Holland, J. Kuti, D. Nogradi, C. Schroeder, and C.-H. 
Wong, Phys. Lett. B {\bf 703}, 348 (2011); 1211.4238; 
Y. Aoki et al., Phys. Rev. D {\bf 86}, 054506 (2012); and
A. Hasenfratz, A. Cheng, G. Petropoulos, and D. Schaich, arXiv:1207.7162. 

\bibitem{dil}
%
Some of the theoretical papers on the mass of a possible light dilaton include
the following (this list focuses on papers with dilaton mass estimates; other
papers concentrate on phenomenological fits of the 126 GeV boson discovered at
the LHC to a dilaton with various assumptions about the dilaton couplings):
K. Yamawaki, M. Bando, and K. Matumoto, Phys. Rev. Lett. {\bf 56}, 1335 (1986);
W. A. Bardeen, C. N. Leung, and S. T. Love, Phys. Rev. Lett. {\bf 56}, 1230
(1986); B. Holdom and J. Terning, Phys. Lett. B {\bf 187}, 357 (1987);
Phys. Lett. B {\bf 200}, 338 (1988); D. D. Dietrich, F. Sannino, and
K. Tuominen, Phys. Rev. D { \bf 72 }, 055001 (2005); W. D. Goldberger,
B. Grinstein, and W. Skiba, Phys. Rev. Lett. {\bf 100}, 111802 (2008);
J.J. Fan, W. D. Goldberger, A. Ross, and W. Skiba, Phys. Rev. D {\bf 79},
035017 (2009); T. Appelquist and Y. Bai, Phys. Rev. D {\bf 82}, 071701 (2010);
D. Elander, C. N\'u\~{n}ez, and M. Piai, Phys. Lett. B {\bf 686}, 64 (2010);
A. Delgado, K. Lane, and A. Martin, Phys. Lett. B {\bf 696}, 482 (2010);
B. Grinstein and P. Uttayarat, JHEP 1107, 038 (2011); M. Hashimoto and
K. Yamawaki, Phys. Rev. D {\bf 83}, 015008 (2011); O. Antipin, M. Mojaza, and
F. Sannino, Phys. Lett. B {\bf 712}, 119 (2012); S. Matsuzaki and K. Yamawaki,
Phys. Rev. D {\bf 85}, 095020 (2012); S. Matsuzaki and K. Yamawaki,
Phys. Rev. D {\bf 86}, 035025, 115004 (2012); L. Anguelova, P. Suranyi, and
L. C. R. Wijewardhana, Nucl. Phys. B {\bf 862}, 671 (2012); R. Lawrance and
M. Piai, arXiv:1207.0427; D. Elander and M. Piai, arXiv:1208.0546; Z. Chacko
and R. K. Mishra, arXiv:1209.3022; B. Bellazini, C. Cs\'aki, J. Hubisz,
J. Serra, and J. Terning, arXiv:1209.3299. Some progress toward the goal of a
lattice measurement of a dilaton mass is reported in Z. Fodor et al.,
arXiv:1211.6164 and in talks by M. Lin at the workshop ``Lattice Meets
Experiment: Beyond the Standard Model'', Univ. of Colorado, Oct. 26-27, 2012
and by J. Kuti, E. Neil, and E. Rinaldi at the workshop on Strongly Coupled
Gauge Theories, SCGT12, Nagoya, Japan, Nov. 4-7, 2012.

\bibitem{casimir}
%
The quadratic Casimir invariant $C_2(R)$ for the
representation $R$ is given by $\sum_{a=1}^{o(G)} \sum_{j=1}^{dim(R)}
[D_R(T_a)]_{ij} [D_R(T_a)]_{jk} = C_2(R)\delta_{ik}$, where $a,b$ are group
indices, $o(G)$ is the order of the group, $T_a$ are the generators of the
associated Lie algebra, and $D_R(T_a)$ is the matrix form of the $T_a$ in the
representation $R$.  The trace invariant $T(R)$ is defined by
$\sum_{i,j=1}^{dim(R)}[D_R(T_a)]_{ij} [D_R(T_b)]_{ji} = T(R)\delta_{ab}$.
In our notation, $C_A \equiv C_2(G)$, and $C_f \equiv C_2(R)$, 
$T_f \equiv T(R)$ for the representation $R$. Thus, e.g.,
for $G={\rm SU}(N_c)$, $C_A=N_c$, and for $R=\fund$, 
$C_2(\fund)=(N_c^2-1)/(2N_c)$, $T(\fund)=1/2$. 

\bibitem{nfintegral}
%
Here and below, when an expression is given for $N_f$ that formally evaluates
to a non-integral real value $\nu$, it is understood implicitly that one infers
an appropriate integral value of $N_f$ from it, either the greatest integer
less than $\nu$, or the smallest integer greater than $\nu$, or the integer
nearest to $\nu$, depending on the context.  Parenthetically, we note that 
although our analysis applies to theories with $N_f$ Dirac fermions, in
the special case of a fermion in the adjoint representation, one may formally 
use the value $N_f=1/2$ to describe a theory with a Majorana fermion.

\bibitem{othergg}
%
See M. Mojaza, C. Pica, T. A. Ryttov, and F. Sannino, Phys. Rev. D {\bf 86}, 
076012 (2012) for studies of other gauge groups. 

\bibitem{disc}
I. M. Gelfand, M. M. Kapranov, and A. V. Zelevinsky, {\it Discriminants, 
Resultants, and Multidimensional Determinants} (Birkh\"auser, Boston, 1994). 

\bibitem{symfun}
J. V. Uspensky, {\it Theory of Equations} (McGraw-Hill, New York, 1948); 
%I. G. Macdonald, {\it Symmetric Functions and Hall Polynomials} 
%(Oxford University Press, Oxford, 2005);

\bibitem{otherpade}
%
The other two Pad\'e approximants with $p+q=3$ are [3,0], which is identical to
$\beta_{4\ell,r}$ itself, and [0,3], which is not of interest here, since it
has no zeros. 

\bibitem{bethke}
%
A recent review is S. Bethke, Eur. Phys. J. C {\bf 64}, 689 (2009).

\bibitem{brodskyschemes}
%
See, e.g., S. J. Brodsky and X.-G. Wu, Phys. Rev. Lett. {\bf 109}, 042002
(2012); Phys. Rev. D { \bf 86}, 014021 (2012); P. M. Stevenson, Nucl. Phys. B
{\bf 868}, 38 (2013); M. Mojaza, S. J. Brodsky, and X.-G. Wu., arXiv:1212.0049
and references therein to the earlier literature.

\bibitem{alt}
%
Alternatively, we could use the weaker assumption that $b_\ell$ with
$\ell=3,...,n$ are such as to preserve the IR zero of $\beta_{2\ell}$ for $N_f
\in I$ up to $n$-loop order and consider the shift in the IR zero due to
increasing $|b_{n+1}|$ from zero. By continuity, the existence of the $n$-loop
IR zero is preserved for small $|b_{n+1}|$.

\bibitem{nsvzbeta}
V. A. Novikov, M. A. Shifman, A. I. Vainshtein, and V. I. Zakharov, Nucl. 
Phys. B {\bf 229}, 381 (1983); Nucl. Phys. B {\bf 277}, 426 (1986).

\bibitem{seiberg}
N. Seiberg, Phys. Rev. D {\bf 49}, 6857 (1994); Nucl. Phys. B {\bf 435},
129 (1995); K. A. Intriligator and N. Seiberg, Nucl. Phys. B {\bf 444}, 125 
(1995). 

\bibitem{b1s}
D. R. T. Jones, Nucl. Phys. B {\bf 87}, 127 (1975).

\bibitem{b2s}
%
M. Machacek and M. Vaughn, Nucl. Phys. B {\bf 222}, 83 (1983);
A. J. Parkes and P. C. West, Phys. Lett. B {\bf 138}, 99 (1984);
Nucl. Phys. B {\bf 256}, 340 (1985);
D. R. T. Jones and L. Mezincescu, Phys. Lett. B {\bf 136}, 242 (1984);
Phys. Lett. B {\bf 138}, 293 (1984).

\bibitem{b3s}
%
R. V. Harlander, D. R. T. Jones, P. Kant, L. Mihaila, and M. Steinhauser,
JHEP 0612, 024 (2006);
R. Harlander, L. Mihaila, and M. Steinhauser (HMS), Eur. Phys. J.
C {\bf 63}, 383 (2009). 

\bibitem{dred}
%
W. Siegel, Phys. Lett. B {\bf 84}, 193 (1979); Phys. Lett. B {\bf 94}, 37 
(1980); a recent discussion is W. St\"ockinger, JHEP 0503, 076 (2005).

\bibitem{rsbeta}
T. A. Ryttov and F. Sannino, Phys. Rev. D {\bf 78}, 065001 (2008); 
C. Pica and F. Sannino, Phys. Rev. D {\bf 83}, 116001 (2011). 

\bibitem{sth}
R. Oehme, Phys. Lett. B {\bf 399}, 67 (1997); 
M. T. Frandsen, T. Pickup, and M. Teper, Phys. Lett. B {\bf 695}, 231 (2011).

\end{thebibliography}
\end{document}